\documentclass[11pt]{article} 
\usepackage{jheppub}

\usepackage{amsmath,amsfonts,amsbsy,amssymb,array,accents,dsfont}
\usepackage{enumerate}
\usepackage[shortlabels]{enumitem}
\usepackage{graphicx} 
\usepackage{subcaption} 
\usepackage{upgreek}
\usepackage{bm}
\usepackage{slashed}

\let\includefigures=\iftrue
\let\useblackboard==\iftrue
\newfam\black

\usepackage[dvipsames]{xcolor}

\definecolor{myblue}{RGB}{85,130,255}%{55, 100, 210}
\definecolor{myred}{RGB}{200, 45, 40}

\PassOptionsToPackage{ 
     colorlinks=true,
     linkcolor=myBlue,
     urlcolor=blue,
     filecolor=black,
     citecolor=myRed,
}{hyperref}

\usepackage{xparse}
\usepackage{xspace}

\NewDocumentCommand\eqn{om}{%
  \IfNoValueTF{#1}
     {\[ #2 \]}
     {\begin{equation}\label{#1} #2  \end{equation} \expandafter\newcommand\csname #1\endcsname{\eqref{#1}\xspace}\ignorespaces}
}
\NewDocumentCommand\eqna{om}{%
  \IfNoValueTF{#1}
    {\begin{align*} #2 \end{align*}}
    {\begin{equation}\label{#1}\begin{split} #2  \end{split}\end{equation} \expandafter\def\csname #1\endcsname{\eqref{#1}\xspace}\ignorespaces}
}

\newcommand{\rcite}{\cite}

\def\that{{\tt t}}
\def\rhat{{\tt r}}

\def\ThetaN{{\Theta^{\vphantom{|}}_{\!N}}}
\def\Mtil{{\widetilde M}}
\def\Jtil{{\widetilde J}}
\def\Ntil{{\widetilde N}}

\def\lapse{{ \sfN}}

\def\varphib{{\boldsymbol\varphi}}

\def\taub{{\boldsymbol\tau}}

\def\sl{\text{sl}}

\def\eps{\epsilon}
\def\vareps{\varepsilon}

\def\ptcl{{\rm ptcl}}
\def\str{{\rm str}}
\def\BH{{\rm BH}}
\def\BTZ{{\rm BTZ}}

\def\sltwo{\ensuremath{SL(2,\bR)}}

\def\mathbi#1{\textbf{\em #1}}
\def\tight#1{\! #1 \!}  % tightens annoying spacing in equations

\def\rbi{{\mathbi r}}
\def\pbi{{\mathbi p}}

\def\({\left(}
\def\){\right)}
\def\[{\left[}
\def\]{\right]}

\def\ie{{i.e.}}
\def\eg{{e.g.}}

\def\btz{{\sst \rm BTZ}}

\def\sfH{{\mathsf H}}
\def\sfI{{\mathsf I}}

\def\sfN{{\mathsf N}}
\def\sfP{{\mathsf P}}

\def\sfc{{\mathsf c}}

\def\sfm{{\mathsf m}}

\DeclareMathSymbol{\medhatsym}{\mathord}{largesymbols}{"62} % basic symbol

\DeclareMathSymbol{\medtildesym}{\mathord}{largesymbols}{"65}% basic symbol

\makeatletter
\newcommand*\rel@kern[1]{\kern#1\dimexpr\macc@kerna}
\newcommand*\widebar[1]{%
  \begingroup
  \def\mathaccent##1##2{%
    \rel@kern{0.8}%
    \overline{\rel@kern{-0.8}\macc@nucleus\rel@kern{0.2}}%
    \rel@kern{-0.2}%
  }%
  \macc@depth\@ne
  \let\math@bgroup\@empty \let\math@egroup\macc@set@skewchar
  \mathsurround\z@ \frozen@everymath{\mathgroup\macc@group\relax}%
  \macc@set@skewchar\relax
  \let\mathaccentV\macc@nested@a
  \macc@nested@a\relax111{#1}%
  \endgroup
}
\makeatother

\def\half{\frac12}

\def\hf{\coeff12}
\def\tr{{\rm Tr}}
\def\Tr{{\rm Tr}}
\def\One{{\hbox{1\kern-1mm l}}}

\def\Im{{\sfI\sfm\,}}

\def\barray{\begin{array}}
\def\earray{\end{array}}
\def\be{\begin{equation}}
\def\ee{\end{equation}}
\def\bea{\begin{eqnarray}}
\def\eea{\end{eqnarray}}
\def\bal{\begin{align}}
\def\eal{\end{align}}

\newcommand{\bR}{{\mathbb R}}
\newcommand{\bS}{{\mathbb S}}
\newcommand{\bT}{{\mathbb T}}

\newcommand{\bZ}{{\mathbb Z}}

\definecolor{cardinal}{rgb}{0.6,0,0}
\definecolor{darkgreen}{rgb}{0,0.4,0}
\definecolor{green}{rgb}{0,0.4,0}
\definecolor{golden}{rgb}{0.92, 0.7, 0}
\definecolor{midnight}{rgb}{0, 0, 0.5}
\definecolor{darkblue}{rgb}{0, 0, 0.7}

\numberwithin{equation}{section}

\mathchardef\mhyphen="2D

\def\cM{\mathcal {M}}  
\def\cP{\mathcal {P}}  
\def\cS{\mathcal {S}}

\def\one{{\hbox{\kern+.5mm 1\kern-.8mm l}}}
\def\zero{{\hbox{0\kern-1.5mm 0}}}

\def\id{\textrm{id}}

\def\id{{1 \kern-.28em {\rm l}}}

\def\journal#1&#2(#3){\unskip, \sl #1\ \bf #2 \rm(19#3) }
\def\andjournal#1&#2(#3){\sl #1~\bf #2 \rm (19#3) }

\def\ie{{\it i.e.}}
\def\eg{{\it e.g.}}

\def\sst{\scriptscriptstyle}

\def\half{\frac12}
\def\hf{{\textstyle\half}}

\def\One{{1\hskip -3pt {\rm l}}}

\catcode`\@=11
\def\slash#1{\mathord{\mathpalette\c@ncel{#1}}}
\def\eps{\epsilon}
\def\vareps{\varepsilon}
\def\underrel#1\over#2{\mathrel{\mathop{\kern\z@#1}\limits_{#2}}}

\catcode`\@=12

\def\exp{{\rm exp}}
\def\ie{{\it i.e.}}
\def\eg{{\it e.g.}}

\title{%\LARGE 
{
The Holar Wind
}}

\author{
Emil J. Martinec
}

\affiliation{
\vskip 0.01cm
Kadanoff Center for Theoretical Physics, Enrico Fermi Institute, and Department of Physics\\ 
University of Chicago,
5640 S. Ellis Ave.,
Chicago IL 60637\\ 
}

\emailAdd{%
e-martinec@uchicago.edu}

\abstract{%
String theory in $AdS_3$ with purely NS-NS fluxes and vanishing RR moduli has a continuum of winding string excitations in radial plane wave states.  BTZ black holes can emit such strings, which then flow out toward the $AdS_3$ boundary as a stream of massive quanta, and form a black hole analogue of the solar wind.
The winding string sector thus provides a decay channel for the black hole to evaporate without having either to couple the system to an external reservoir or to match the $AdS_3$ throat onto an asymptotically flat region.
We compute the emission amplitude of this ``holar wind" in the semi-classical approximation, and consider the associated version of the black hole information paradox.
}

\dedicated{\begin{center}\it 
Dedicated to the memory of Eugene N. Parker %(1927-2022)
\end{center}}

\begin{document}
\hypersetup{pageanchor=false}
\begin{titlepage}
\maketitle
\thispagestyle{empty}
\end{titlepage}
\hypersetup{pageanchor=true}
\pagenumbering{arabic}

%\toc
\thispagestyle{empty}

%\vskip 1cm
%\hrule

%%%%%%%%%%%%%%%%%%%%%%%%%%%%%%%%%%%%%%%%%%%%%%%%%%%%%%%%%%%%%%%%
%%%%%%%%%%%%%%%%%%%%%%%%%%%%%%%%%%%%%%%%%%%%%%%%%%%%%%%%%%%%%%%%

%%%%%%%%%%%%%%%%%%%%%%%%%%%%%%%%%%%%%%%%%%%%%%%%%%%%%%%%%%%%%%%%
%%%%%%%%%%%%%%%%%%%%%%%%%%%%%%%%%%%%%%%%%%%%%%%%%%%%%%%%%%%%%%%%

%\begin{enumerate}[start=1,
%    labelindent=\parindent,
%    leftmargin =2.5\parindent,
%    label=(WII-\arabic*)]
%\item
%\label{WII-1}
%stuff
%\item
%\label{WII-2}
%more stuff
%\end{enumerate}

%Later on I want to refer to \ref{WII-1}

%%%%%%%%%%%%%%%%%%%%%%%%%%%%%%%%%%%%%%%%%%%%%%%%%%%%%%%%%%%%%%%%
%%%%%%%%%%%%%%%%%%%%%%%%%%%%%%%%%%%%%%%%%%%%%%%%%%%%%%%%%%%%%%%%

\section{Introduction} 
\label{sec:intro}

The $AdS_3/{\it CFT}_{\!2}$ correspondence has proven a particularly fruitful example of holography.  It arises for example in a particular decoupling limit of NS5-brane dynamics, in which $n_5$ coincident fivebranes are wrapped around a compactification $\bS^1\times\cM$, where $\cM=\bT^4$ or $K3$ has a size of order the string scale.  In addition, one binds $n_1$ fundamental (F1) strings to the fivebranes which also wrap the $\bS^1$; then the near-source geometry is $AdS_3\times \bS^3\times \cM$, with the $\bS^1$ becoming the azimuthal direction of~$AdS_3$.

One of the attractive features of this corner of string theory is the availability of a solvable worldsheet dynamics~\rcite{Giveon:1998ns,Kutasov:1999xu,Maldacena:2000hw,Maldacena:2001km,Giveon:1999px,Giveon:1999tq}, in which the $AdS_3$ dynamics is governed by $\sltwo$ current algebra.  
The electric $H_3$ flux sourced by the background strings supplies the Wess-Zumino term of the $\sltwo$ WZW model.  For a string propagating on this background, the Lorentz force of the background electric $H_3$ field counterbalances the string tension, such that there exists a continuum of ``long string'' states in which strings wrap the azimuthal direction yet can expand radially outward toward the $AdS_3$ conformal boundary at finite cost in energy.%
\footnote{The worldsheet formalism describes the codimension four locus of the moduli space in which all RR moduli are set to zero.  Turning on any of these RR moduli alters the balance of tension and Lorentz force, and prevents these winding strings from reaching the conformal boundary; the long string spectrum develops a gap~\rcite{Seiberg:1999xz}.  Our interest is in exploiting the long string continuum, so we will work at this ``singular'' locus of the CFT moduli space.}
Because it is an intrinsic feature of the worldsheet affine $\sltwo$ representation theory, a wide variety of $AdS_3$ string compactifications exhibit this long string continuum, including%
~\rcite{%
Elitzur:1998mm,
deBoer:1998gyt,
Kutasov:1998zh,
Giveon:1999jg,
Giveon:1999zm,
Argurio:2000tg,
Argurio:2000tb,
Argurio:2000xm}.

Ordinarily in the AdS/CFT correspondence, the observables are CFT correlators, computed in the worldsheet theory via the perturbative string version of Witten diagrams~\rcite{Kutasov:1999xu,Maldacena:2001km}.
The long string continuum instead has an S-matrix.  
Long strings come in from and go out to timelike infinity at the spatial boundary.  In particular, in the black hole sector of the theory, the S-matrix consists of long strings being absorbed by a BTZ black hole, and a tiny amplitude for long strings to be emitted as Hawking radiation.

In this work, we evaluate the amplitude for BTZ black holes to radiate long strings.  We begin in section~\ref{sec:classical} with a review of the classical dynamics of particles~\rcite{Farina:1993xw,Cruz:1994ir} and strings%
~\rcite{%
Natsuume:1996ij,
Hemming:2001we,
Hemming:2002kd,
Troost:2002wk,
Ashok:2021ffx,
Nippanikar:2021skr,
Ashok:2022vdz} 
in the BTZ geometry.
Then in section~\ref{sec:hawking} we review the computation of the emission of Hawking quanta from BTZ black holes, and extend the computation to the winding string sector.  There is a tiny amplitude for a long string to be emitted with sufficient radial momentum to put it in the long string continuum, thus generating a black hole S-matrix within $AdS_3$.   

The strings being absorbed and emitted carry the fundamental string winding charge which determines the $AdS_3$ radius in Planck units.  The black hole is thus effectively living in a grand canonical ensemble where the $AdS_3$ cosmological constant is a thermodynamic variable that is allowed to fluctuate~\rcite{Caldarelli:1999xj,Kastor:2009wy,Cvetic:2010jb,Wang:2006eb,Dolan:2010zz,Dolan:2010ha}.  We take this fact into account in the process of evaluating the emission probability.  All told, we find that the probability of a winding string to tunnel out of the black hole is determined by the change in the black hole entropy in this ensemble
\be
\Gamma \sim \exp\big[ \Delta S_\BH \big]  ~,
\ee
in line with the result for particle emission~\rcite{Keski-Vakkuri:1996wom,Massar:1999wg,Parikh:1999mf} and expectations from thermodynamics.

In section~\ref{sec:unitarity}, we consider the consequences of this result for the black hole information problem in the context of the AdS/CFT correspondence.  BTZ black holes can relax back to extremality by emitting long strings, and various proposals for recovering information from the radiation can be examined.
Here the issue is the following: Is the string being radiated a scrambled version of the ones that made up the initial BTZ black hole, or is it a new one drawn from vacuum fluctuations near the horizon, which has nothing to do with the ones making up the brane bound state?  Unitarity of the evolution requires the former, while the Hawking process that generates black hole radiation in effective field theory suggests the latter.  We compare and contrast various proposals for maintaining the unitarity of black hole radiance in light of our results, including the {\it fuzzball scenario} (see~\rcite{Bena:2022rna} for an overview), and the {\it island scenario} (see~\rcite{Almheiri:2020cfm} for a review).

%%%%%%%%%%%%%%%%%%%%%%%%%%%%%%%%%%%%%%%%%%%%%%%%%%%%%%%%%%%%%%%%
%%%%%%%%%%%%%%%%%%%%%%%%%%%%%%%%%%%%%%%%%%%%%%%%%%%%%%%%%%%%%%%%

\section{Classical BTZ geometry, geodesics, and winding strings}
\label{sec:classical}

The BTZ solution in dimensionless static coordinates is
\begin{align}
\label{BTZmetric}
ds^2 = \ell^2\Big[ - f(r) dt^2 +\frac{dr^2}{f(r)} + r^2\Big( d\phi - \frac{r_+r_-}{ r^2}\, dt\Big)^2 \Big]
~~,~~~~
f(r) = \frac{(r^2-r_+^2)(r^2-r_-^2)}{ r^2}
\end{align}
The $AdS_3$ radius $\ell$ is related to the 3d Planck scale $G$ via
\be
\label{centchg}
c = \frac{3\ell}{2G} \equiv 6N = 6 \, k n_1  ~,
\ee
where $c$ is the central charge of the dual $\it CFT_2$ and $n_1$ is the F1 charge of the background.
For concreteness, we can consider the F1-NS5 system compactified on $\bT^4$ or $K3$, for which $N=n_1n_5$ and $k=n_5$ in terms of the numbers of strings and fivebranes in the background; however our results hold for any $AdS_3$ string theory with purely NS flux.

\begin{figure}[ht]
\centering
\includegraphics[width=0.4\textwidth]{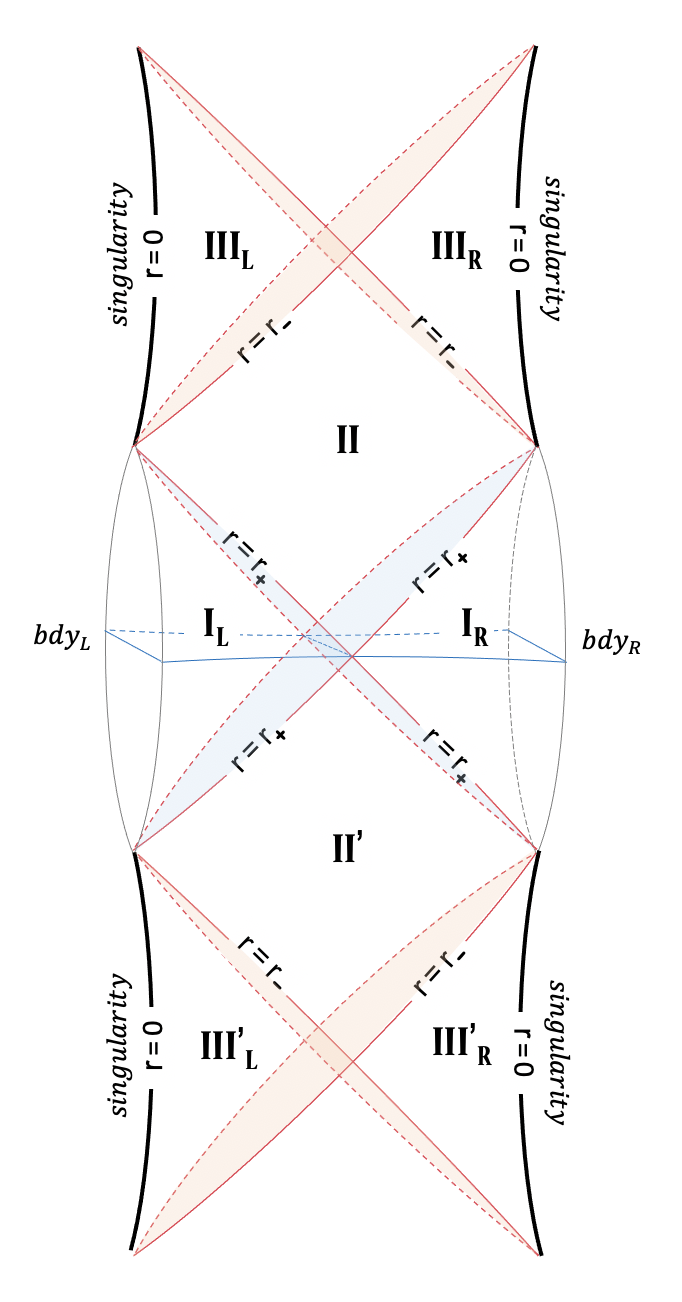}
\caption{\it Penrose diagram of a rotating BTZ black hole.  The back and front are identified to make the $\phi$ coordinate periodic.  The outer (inner) horizon is shaded in blue (red); the surface of time symmetry $t=0$ is outlined with a thin blue line.}
\label{fig:Penrose3d}
\end{figure}

The inner and outer horizon radii $r_\pm$ are related to the mass and angular momentum of the black hole via
\be
\label{MJrpm}
\ell M = \frac{\ell (r_+^2+r_-^2)}{8G} = \frac N2 \big( {r_+^2 + r_-^2} \big)
~~,~~~~
J = \frac{\ell \, r_+ r_-  }{4G} = N\, r_+ r_- ~.
\ee 
The Penrose diagram for this geometry is depicted in figure~\ref{fig:Penrose3d}.

We collect here for later use some of the thermodynamic properties of BTZ black holes.  The entropy, temperature and angular potential
\be
\label{SandT}
S_\btz = \frac{A}{4G} = \frac{2\pi\ell r_+}{4G} = 2\pi N r_+
~~,~~~~
T = \frac{\kappa_+}{2\pi} = \frac{r_+^2 - r_-^2}{2\pi \ell r_+}  
~~,~~~~
\Omega_+ = \frac{r_-}{\ell r_+} 
\ee
specify the first law of black hole thermodynamics
\be
\label{firstlaw}
dM = T dS + \Omega_+ dJ
\ee
with as usual $\Omega_+$ being the angular velocity of the outer horizon, and the temperature given by the surface gravity $\kappa_+$ at the outer horizon.  The entropy~\eqref{SandT} can be recast as that of a (dual) 2d CFT
\begin{align}
\label{Scft}
S_\btz = 2\pi\Big(\sqrt{\vphantom{\bar L_0} NL_0} \,+ \sqrt{N\bar L_0} \; \Big) ~,
\end{align}
where
\begin{align}
\begin{split}
\label{Lzero}
L_0 = \frac{\ell (r_+ \tight+ r_-)^2}{16G} = N\frac{(r_+ \tight+ r_-)^2}{4} = \half\big( \ell M+J \big)
\\[.2cm]
\bar L_0 = \frac{\ell (r_+ - r_-)^2}{16G} = N\frac{(r_+ - r_-)^2}{4} = \half\big( \ell M-J \big)
\end{split}
\end{align}
are the left and right conformal dimensions of the state.

%
%%%%%%%%%%%%%%%%%%
%%%%%%%%%%%%%%%%%%
%

The BTZ geometry has constant negative curvature, with $\ell$ the curvature radius; indeed it can be thought of a quotient of the $\sltwo$ group manifold by the discrete group action
\be
\label{btzquotient}
g \mapsto e^{\pi(r_+-r_-)\sigma_3} \,g\, e^{\pi(r_++r_-)\sigma_3}  ~.
\ee
As such, ithe worldsheet action of a string propagating in this background is given by
\be
\label{Sstr}
\cS_\str = \frac{1}{4\pi \alpha'} \int\!d^2\xi \sqrt{\gamma}\Big(\gamma^{ab} \partial_a x^\mu\partial_bx^\nu G_{\mu\nu}(x)
+ \epsilon^{ab} \partial_ax^\mu\partial_bx^\nu B_{\mu\nu}(x) \Big)
\ee
where the metric~\eqref{BTZmetric} is locally that of the $\sltwo$ WZW model at level 
\be
\label{level}
k=\ell^2/\alpha'
\ee
(we will work in conventions where $\alpha'=1$),
and the Wess-Zumino term corresponds to an $H_3$ flux given by the covariantly constant volume form 
$\Tr\big[(g^{-1}dg)^3\big]$, 
so that
\be
\label{Bfield}
B_{t\phi} =  \ell^2\big( r^2-\sfc\big) ~.
\ee
We have included the possibility of an integration constant $\sfc$ in determining $B$ from $H_3$, which can be adjusted by a gauge transformation.  

A variety of choices of $\sfc$ have been suggested in the literature.  Starting from the $B$-field $B=\ell^2r^2$ in global $AdS_3$ and performing the identification~\eqref{btzquotient} leads to $\sfc=r_-^2$~\rcite{Hemming:2001we,Hemming:2002kd};
smoothness of the Euclidean solution at the horizon gives $\sfc=r_+^2$~\rcite{Ashok:2021ffx} (see also~\rcite{Rangamani:2007fz});
the choice $\sfc=r_+^2+r_-^2$ was suggested in~\rcite{Nippanikar:2021skr} on the basis of symmetry considerations; and
an analysis of the energetics of winding strings suggests $\sfc=0$~\rcite{Ashok:2021ffx}.  The choice of $\sfc$ doesn't affect classical particle or string dynamics, but will affect black hole thermodynamics for which the $B$-field at the horizon will be the chemical potential for the emission of winding strings.  Below we will argue that the correct choice is $\sfc=0$.

In terms of $\sltwo$ group elements, the Euler angle parametrization
\be
\label{euler}
g = e^{(r_+-r_-)(t+\phi){\sigma_3}/{2}}\, e^{\rho \sigma_1}\, e^{-(r_++r_-)(t-\phi)\sigma_3/2}  
\ee
yields the metric~\eqref{BTZmetric} as the standard $\sltwo$ bi-invariant metric $ds^2 \tight= -\half{\ell^2}\, \tr[\,dg \, dg^{-1}]$ on the group manifold in region I of the Penrose diagram (and its copies under analytic continuation), with the radial coordinate $r$ related to $\rho$ via
\be
\cosh^2\rho = \frac{r^2-r_-^2}{r_+^2-r_-^2}  ~.
\ee
Another Euler angle parametrization 
\be
g = e^{(r_+-r_-)(t+\phi){\sigma_3}/{2}}\, e^{i(\tilde\rho-\pi/2) \sigma_2}\, e^{-(r_++r_-)(t-\phi)\sigma_3/2} 
\ee
describes region II (and its copies under analytic continuation), with 
\be
\sin^2\tilde\rho = \frac{r^2-r_-^2}{r_+^2-r_-^2} ~.
\ee
A similar parametrization describes region III and its iterates.   
The identification~\eqref{btzquotient} is simply the periodic identification $\phi\sim\phi+2\pi$.

These three regions and their universal cover in $\phi$ do not fill out the entire $\sltwo$ group manifold; the rest is obtained by the analytic extension to negative values of $r^2$, resulting in a region of closed timelike curves~\rcite{Hemming:2002kd,Nippanikar:2021skr} (since $\phi$ is periodic and timelike there) which is usually ignored as unphysical.  How much of this Penrose diagram is relevant to string theory remains an open question.  The region of negative $r^2$ is presumably an artifact of the analytic continuation.  In a situation where the black hole is formed by collapse, regions $\rm I_L$, $\rm II'$ and further to the past are excised and replaced by the collapsing source geometry.  Furthermore, it has been argued~\rcite{Marolf:2011dj} that the region inside the outgoing portion of the inner horizon (the null surface separating regions $\rm III_L$ and $\rm II$) is unstable to perturbations, forming an exponentially growing shockwave along this portion of $r=r_-$.  One may suppose that in string theory, this shockwave and its extreme tidal forces are resolved by the gas of branes that the shockwave devolves into under tidal disruption~\rcite{Horowitz:1990sr,Martinec:2020cml}.   Finally, the fuzzball paradigm posits that even region II is supplanted by a deconfined, non-geometric phase of brane matter that supplies the underlying microstates responsible for the black hole entropy~\rcite{Mathur:2005zp}.  We will have more to say about this proposal below.

%%%%%%%%%%%%%%%%%%%%%%%%%%%%%%%%
%%%%%%%%%%%%%%%%%%%%%%%%%%%%%%%%

\subsection{Geodesics in BTZ}
\label{sec:geodesics}

\begin{figure}[ht]
\centering
  \begin{subfigure}[b]{0.3\textwidth}
  \hskip 0cm
    \includegraphics[width=\textwidth]{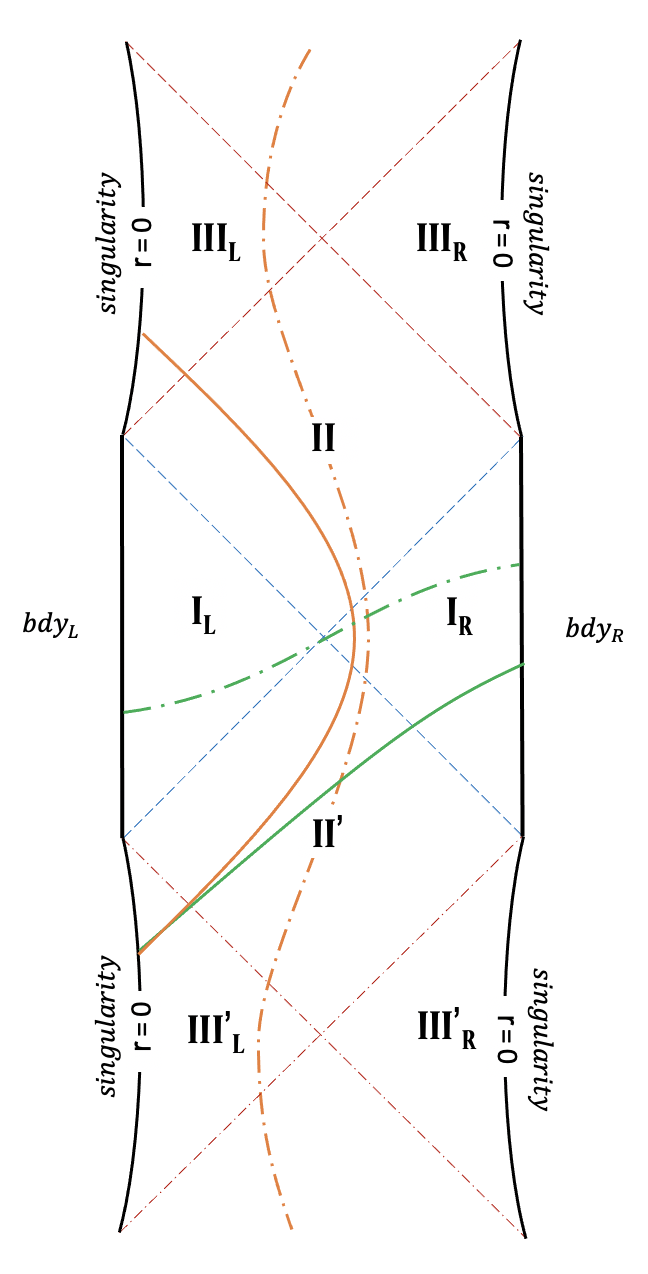}
    \caption{ }
    \label{fig:geodesics}
  \end{subfigure}
\qquad\qquad\qquad
  \begin{subfigure}[b]{0.3\textwidth}
      \hskip 0cm
    \includegraphics[width=\textwidth]{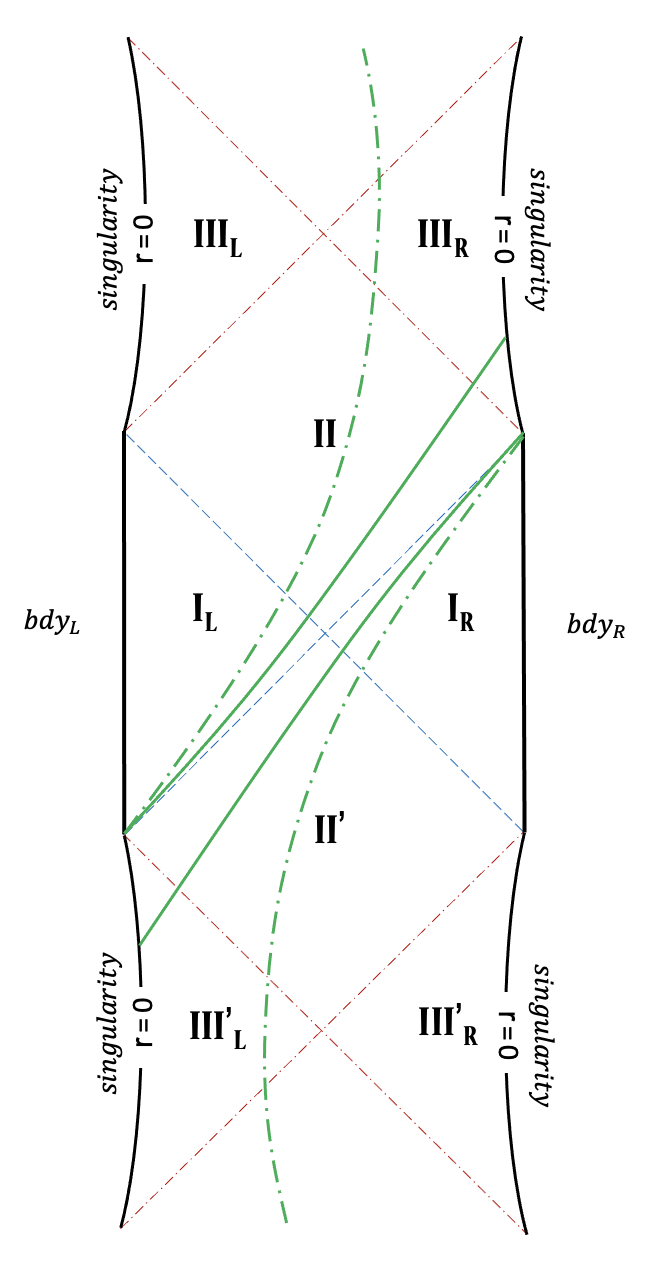}
    \caption{ }
    \label{fig:windingtrajectories}
  \end{subfigure}
\caption{\it 
(a) Classical pointlike trajectories are geodesics, whose projection onto the $r$-$t$ plane is depicted.  Timelike trajectories are in orange, spacelike in green; low angular momentum are the solid curves, high angular momentum are the dashed curves.  Null trajectories are the limit $\mu\to 0$ of the timelike trajectories; note that $r_{\rm max}\to\infty$ in this limit~-- the trajectory reaches the boundary and reflects back.
(b) Classical unbound wound string trajectories are spectral flows of the spacelike geodesics of the left figure.  Solid curves depict low angular momentum, dashed curves high angular momentum.  Two of each, time reversals of one another, are sketched.  Hawking pair production of strings involves tunneling from the backward trajectory inside the horizon to the forward trajectory outside the horizon.  There are also bound winding string trajectories, which look like the timelike particle trajectories of figure (a).
}
\label{fig:Ellipses}
\end{figure}

The dimensional reduction of~\eqref{Sstr} to the worldline of a particle
\be
\label{Sptcl}
\cS_{\rm ptcl} = \half \int \!d\xi \sqrt{\gamma}\Big[ \gamma^{-1} G_{\mu\nu}(x) \partial_\xi x^\mu \partial_\xi x^\nu - \ell^2 \mu \Big]
\ee
governs geodesic motion on the BTZ geometry, where $\gamma$ is the worldline metric.  Timelike geodesics correspond to $\mu>0$, null geodesics have $\mu=0$, and spacelike geodesics result when $\mu<0$.
The geodesics are completely determined by conservation laws.  The cyclic coordinates $t,\phi$ in this action correspond to Killing vectors $\partial_t,\partial_\phi$ of the geometry, and associated conserved momenta
\begin{align}
\begin{split}
\label{ptclp}
E&\equiv -p_t/k = \big(r^2 - r_+^2 - r_-^2\big)\dot t +   r_+r_- \dot\phi 
\\[.2cm]
L &\equiv p_\phi/k =  \big(r^2\dot\phi - r_+r_- \dot t \big)  
\end{split}
\end{align}
where overdots denote $\partial_\xi$.  
The worldline metric $\gamma$ enforces the Hamiltonian constraint
\be
\label{Hamconst}
\sfH = G_{\mu\nu} \dot x^\mu \dot x^\nu + \ell^2 \mu = 0  ~.
\ee
These equations can be integrated to find the radial motion~\rcite{Cruz:1994ir,Hemming:2001we,Troost:2002wk,Ashok:2021ffx,Nippanikar:2021skr}
\be
\label{rsoln}
r^2 = 
\begin{cases}
~\;\,\frac{1}{2\mu}\Big( \alpha + \sqrt{\beta}\, \sin 2\sqrt{\mu}\, \xi \Big)  & \quad \mu>0
\\[.2cm]
-\frac{1}{2\mu}\Big( \alpha + \sqrt{-\beta}\, \sinh 2\sqrt{-\mu}\,\xi \Big) & \quad \mu<0~~,~~ \beta<0
\\[.2cm]
~\;\,\frac{1}{2\mu}\Big( \alpha - \sqrt{\beta}\, \cosh 2\sqrt{-\mu}\,\xi \Big) & \quad \mu<0~~,~~ \beta>0
\end{cases}
\ee
where
\begin{align}
\begin{split}
\alpha &= E^2 - L^2 - \mu (r_+^2 \tight+ r_-^2)
\\[.2cm]
\beta &= \Big( (E\tight-L)^2 + \mu(r_+ \tight- r_-)^2 \Big)\Big( (E\tight+L)^2 + \mu(r_+ \tight+r_-)^2 \Big) ~.
\end{split}
\end{align}
One can also solve for $t(r)$ and $\phi(r)$, with the results~\rcite{Cruz:1994ir}
\begin{align}
t &= \pm \frac{1}{2(r_+^2\tight-r_-^2)}\Big[ r_+ f\big(r^2\tight-r_+^2,B_+,C_+\big) - r_- f\big(r^2-r_-^2,B_-,C_-\big) \Big] + t_0
\\[.2cm]
\phi &= \pm \frac{1}{2(r_+^2\tight-r_-^2)}\Big[ r_- f\big(r^2\tight-r_+^2,B_+,C_+\big) - r_+ f\big(r^2-r_-^2,B_-,C_-\big) \Big] + \phi_0
\\[.2cm]
f(x,&B,C) = \log\Big[\frac{Cx}{2C\sqrt{-\mu x^2+Bx+C^2}\, +2C^2+Bx}\Big]
\\[.2cm]
B_\pm &= E^2\tight-L^2 + (r_\pm^2\tight-r_\mp^2)
~~,~~~~
C_\pm = E r_\pm \tight- L r_\mp
\end{align}
These geodesics are sketched in figure~\ref{fig:geodesics}.

%%%%%%%%%%%%%%%%%%
%%%%%%%%%%%%%%%%%%

The timelike geodesics emerge from the past horizon, reach a maximum radius
\be
r_{\rm max} = \frac{\alpha+\sqrt{\beta}}{2\mu}
\ee
and return to the black hole interior through the future horizon.  For small angular momentum, the trajectory starts and ends at the singularity at $r=0$ (in different copies of region III).  For sufficiently large angular momentum, $\alpha>\sqrt{\beta}$ the angular momentum keeps the motion away from the singularity, and the motion oscillates between the maximum radius $r_{\rm max}>r_+$ and a minimum radius $r_{\rm min}<r_-$.

Spacelike geodesics reach the asymptotic boundary $r\to\infty$ at finite time, and again hit the singularity at $r=0$ for small angular momentum, while for large enough angular momentum they reach a minimum radius $r_{\rm min}<r_-$ and then bounce out into another copy of the asymptotic region~I.

%%%%%%%%%%%%%%%%%%%%%%%%%%%%%%%%
%%%%%%%%%%%%%%%%%%%%%%%%%%%%%%%%

\subsection{Winding strings}
\label{sec:winding}

The center of mass of unwound strings travels along the above geodesics.  Winding strings are obtained from these geodesic trajectories via a {\it spectral flow} operation~\rcite{Maldacena:2000hw,Hemming:2001we,Troost:2002wk,Ashok:2021ffx,Nippanikar:2021skr}
\begin{align}
\begin{split}
\label{specflow}
t(\xi_0,\xi_1) \tight= t_{\ptcl}(\xi_0)\tight+w\xi_0
~~,~~~~
\phi(\xi_0,\xi_1) \tight= \phi_{\ptcl}(\xi_0)\tight+w\xi_1
~~,~~~~
r(\xi_0) \tight= r_{\ptcl}(\xi_0)  ~.
\end{split}
\end{align}
In the Euler angle parametrization~\eqref{euler}, spectral flow amounts to
\be
g \mapsto e^{(r_++r_-)w(\xi_0+\xi_1)\sigma_3/2} \, g \, e^{-(r_+-r_-)w(\xi_0-\xi_1)\sigma_3/2}  ~.
\ee
These wound strings lie in the twisted sectors of the $\bZ$ orbifold~\eqref{btzquotient}.

The currents corresponding to left and right translations along $\sigma_3$ are conserved:
\begin{align}
\begin{split}
J_\str &= -\frac k2\tr\big[g\partial_+ g^{-1} \, \sigma_3\big]  = \frac{k}{r_+ \tight+ r_-}\Big( \big( r^2\tight- r_+^2\tight- r_-^2\tight- r_+r_-\big)\partial_+t + \big(r^2\tight+ r_+r_-\big)\partial_+\phi \Big)
\\[.2cm]
\bar J_\str &= -\frac k2\tr\big[g^{-1}\partial_- g \, \sigma_3\big] = \frac{k}{r_+ \tight- r_-}\Big( \big( r^2\tight- r_+^2\tight- r_-^2 \tight+ r_+r_-\big)\partial_-t - \big(r^2\tight- r_+r_-\big)\partial_-\phi \Big)  ~.
\end{split}
\end{align}
For the geodesics of section~\ref{sec:geodesics} one has
\be
J_\ptcl = \frac k2\frac{E+L}{r_+\tight+ r_-}
~~,~~~~
\bar J_\ptcl = \frac k2\frac{E-L}{r_+\tight- r_-}
~~,~~~~
T_\ptcl = \bar T_\ptcl = -\frac12 k\mu = -\frac{j(j-1)}{k}  ~,
\ee
noting in the last equality that the Hamiltonian is the $\sltwo$ quadratic Casimir.
Spectral flow implements the transformations
\begin{align}
\begin{split}
J_\str &= J_\ptcl - (r_+\tight+ r_-)\frac{kw}{2}
~~,~~~~
\bar J_\str = \bar J_\ptcl - (r_+\tight- r_-)\frac{kw}{2}
\\[.2cm]
T &= T_\ptcl - w(r_+\tight+ r_-) J_\ptcl + \frac k4(r_+\tight+ r_-)^2 w^2
\\[.2cm]
\bar T &= \bar T_\ptcl - w(r_+\tight- r_-) \bar J_\ptcl + \frac k4(r_+\tight- r_-)^2 w^2  ~.
\end{split}
\end{align}
The string energy and angular momentum are the vector and axial current zero modes
\begin{align}
\begin{split}
E_\str &= (r_+\tight+ r_-)J_\str + (r_+\tight- r_-)\bar J_\str   
\\[.2cm]
L_\str &=  (r_+\tight+ r_-)J_\str - (r_+\tight- r_-)\bar J_\str  + kw\, r_+ r_-
\end{split}
\end{align}
where in addition there is a shift in the relation between the axial current and the spacetime angular momentum charge, as noted in~\rcite{Natsuume:1996ij,Hemming:2001we} (see also~\rcite{Ashok:2021ffx}).  This shift ensures that the winding spectrum is level-matched, $L_0-\bar L_0\in\bZ$, if the unflowed state is level-matched.
Note also that we have rescaled the string energy $E_\str$ and angular momentum $L_\str$ by a factor of $k$ relative to their particle counterparts $E,L$ to agree with standard normalizations of the $AdS_3$ worldsheet theory.
Solving the string Virasoro constraints, the energy and spin of the string in the BTZ geometry are coupled to the conformal dimensions $h,\bar h$ of the remaining worldsheet CFT, leading to the mass shell conditions for nonzero winding
\begin{align}
\begin{split}
\label{EstrLstr}
E_\str &= \frac1{w}\Big( {h+\bar h + N_L+ N_R} -\frac{2j(j-1)}k \Big) -\frac{kw}2 \big(r_+^2+r_-^2 \big) 
\\[.2cm]
L_\str &= \frac 1w \Big( h - \bar h +  N_L -  N_R \Big)  
\end{split}
\end{align}
where we also allow for the possibility of $AdS_3$ oscillator excitations $N_L, N_R$, subject to the usual BRST constraints of the worldsheet theory.

The last term in the string energy looks a bit peculiar; naively one might think that the system energy is decreased when a winding string in its ground state is introduced into the system.  However, if we consider the change in energy of the black hole~\eqref{MJrpm} when some amount of string winding $\delta N = kw$ is extracted to make a string of winding $w$ outside the black hole, we have
\be
\label{deltaM}
\delta(\ell M_\BH) = -\frac{kw}{2}\big( r_+^2+r_-^2)
\ee
and this matches the last term in $E_\str$, equation~\eqref{EstrLstr}.  We thus interpret the result of the spectral flow operation as changing the system by extracting some F1 winding charge $w$ from the black hole to make a smaller black hole plus a string carrying winding $w$, with that string carrying energy and angular momentum above threshold
\be
\label{deltaEL}
\delta E = \frac1{w}\Big( {h+\bar h + N_L+ N_R} -\frac{2j(j-1)}k \Big)
~~,~~~~
\delta L = \frac 1w \Big( h - \bar h +  N_L -  N_R \Big)  ~.
\ee
Note that the same logic applies to the winding sector continuum in global $AdS_3$.  This background is the analytic continuation of the BTZ black hole to $r_+^2\to -1, r_-\to 0$, so the background has $\ell M=-N/2$.  The last term in the string energy is~\eqref{EstrLstr} is now positive, and usually thought of as the threshold of the continuum in the sector with winding $w$, with the other terms representing the energy of excitations on the long string.  In the present interpretation, we think of this threshold energy as that required to extract a winding string from the background~-- the NS sector vacuum of the theory is at an energy $L_0 +\bar L_0 = -c/12 = -N/2$, and removing a winding $kw$ from the background {\it raises} its energy by $kw/2$ (the opposite direction as compared to~\eqref{deltaM}).

Bound winding strings are the spectral flow of timelike geodesics with $j\in\bR$, and like particle states, emerge from the past horizon and fall back through the future horizon $r=r_+$.  For low angular momentum, they again hit the singularity; while for high enough angular momentum, they bounce between $0<r_{\rm min}<r_-$ and $r_{\rm max}>r_+$.  

In addition, there is a continuum of unbound winding strings which are the spectral flow of spacelike geodesics, with $j+\half+is$, $s\in\bR$; the parameter $s$ is the radial momentum of the string (conjugate to the coordinate $\rho$).  
While the particle geodesic is spacelike, the winding string solution derived from it by spectral flow has the property that each point along the string has a timelike trajectory.  In particular, it reaches the asymptotic boundary $r\to\infty$ at $t\to\pm\infty$.  For the particle geodesic the $t,\phi$ motions freeze out for large $r$ due to the conservation laws~\eqref{ptclp}, so that asymptotically the geodesic moves radially but not in $t,\phi$.  On the other hand, the winding string motion arising from~\eqref{specflow} asymptotically travels both radially at the rate $s$ (in $\rho\sim \log(r)$) and forward/backward in $t$ at the rate $|w|$, and so both go to infinity together.  The sign of $s$ has been chosen so that the trajectory is radially inbound (outbound) for large negative (positive) $\xi_0$, and the sign of $w$ determines whether the trajectory is past (future) directed (which in the quantum theory translates into whether we are describing strings or anti-strings).

%%%%%%%%%%%%%%%%%%%%%%%%%%%%%%%%%%%%%%%%%%%%%%%%%%%%%%%%%%%%%%%%
%%%%%%%%%%%%%%%%%%%%%%%%%%%%%%%%%%%%%%%%%%%%%%%%%%%%%%%%%%%%%%%%

\section{The Hawking process}
\label{sec:hawking}

Hawking radiation arises in the quantization of fields in a black hole background.  After initial formation of the black hole, the near-horizon region settles down to the vacuum state (as seen by inertial frames of reference) near the horizon, since no excitations of ordinary matter can remain there.  This vacuum state, when decomposed into field modes of asymptotic inertial observers, contains a flux of outgoing (Hawking) radiation.

\begin{figure}[ht]
\centering
\includegraphics[width=0.4\textwidth]{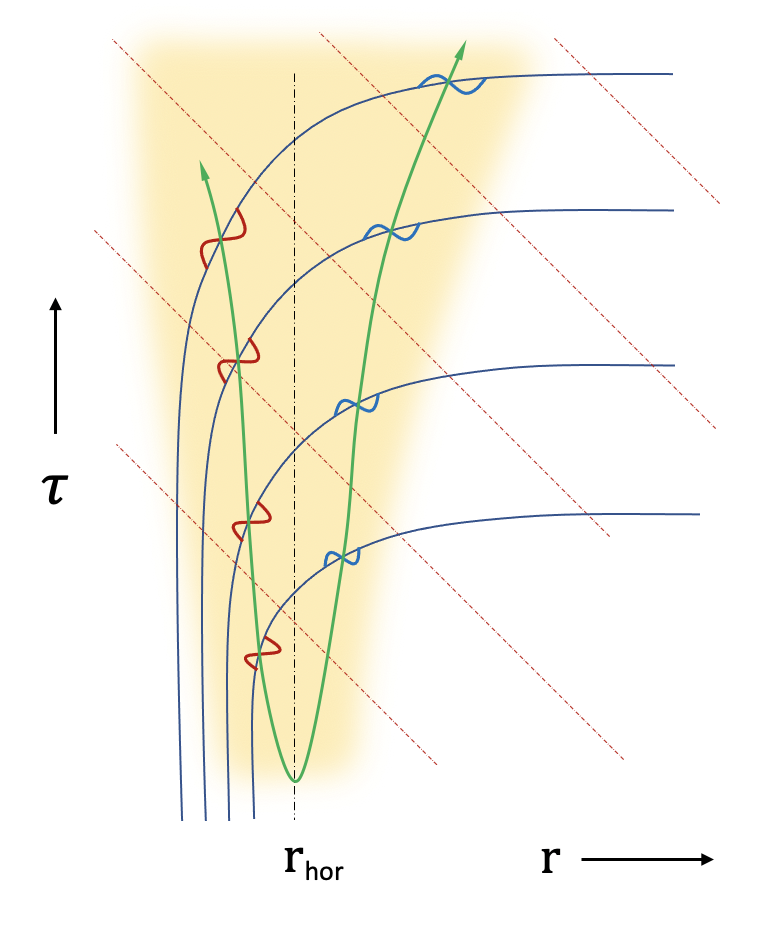}
\vskip -.3cm
\caption{\it Pair creation of Hawking quanta, in a ``nice-slicing" of the near-horizon geometry in 
Painlev\'e-Gullstrand 
%ingoing Eddington-Finkelstein
coordinates.  
The pairs start off as virtual excitations near the horizon but gradually separate.  When the Hawking quanta get sufficiently far from the horizon, they become ordinary particles traveling a semi-classical trajectory of the sort depicted in figure~\ref{fig:geodesics}.}
\label{fig:hawkrad}
\end{figure}

This situation is often thought of in terms of particle creation near the horizon.  Vacuum fluctuations generate virtual particle-antiparticle pairs, one member of which falls into the black hole, with the other escaping the black hole as a quantum of Hawking radiation.  This process is depicted in figure~\ref{fig:hawkrad}.

In appropriate coordinates near the horizon, such as Eddington-Finkelstein or Painlev\'e-Gullstrand coordinates, the particle modes near the horizon have increasingly short wavelength, and a WKB approximation to the wave dynamics becomes appropriate, and except precisely at the horizon follows the geodesics computed above, generating a phase in the quantum amplitude.  The pair creation probability comes from a non-classical part of the trajectory near the horizon, which can be computed as a WKB tunneling amplitude, also governed by the classical actions~\eqref{Sptcl}, \eqref{Sstr}.

%
%%%%%%%%%%%%%%%%%%
%%%%%%%%%%%%%%%%%%
%

When massive particles are created by this process in an AdS black hole background, they don't get very far before reaching a maximum radius, and then they fall back into the black hole.  Massless quanta will reach the AdS boundary, reflect off it, and be focussed back onto the black hole and re-absorbed.  Particles in AdS black hole backgrounds don't have an S-matrix; instead one probes them using operators inserted along the timelike conformal boundary.

On the other hand, we have seen above that winding strings have a continuum of ingoing and outgoing states above some threshold.  Thus we expect a slow but steady flux of outgoing wound strings to gradually cause a BTZ black hole to decay.  Granted, the emission probability is highly exponentially suppressed due to the tiny tunneling amplitude, but it is nonzero~-- we just have to be patient.  To see that the flux of wound strings is no different in terms of its production than the creation of Hawking particles, let us initially consider the near-extremal limit and the resulting $AdS_2$ region near the horizon.  

%%%%%%%%%%%%%%%%%%%%%%%%%%%%%%%%
%%%%%%%%%%%%%%%%%%%%%%%%%%%%%%%%

\subsection{Near-extremal regime and reduction to JT gravity}
\label{sec:JTlimit}

In the near-extremal limit $r_+ - r_-=\varepsilon$, the size of the $\phi$ circle of the BTZ black hole stabilizes near the horizon at $r\sim r_+$, and the geometry has this circle fibered over $AdS_2$.  Defining
\be
\label{AdS2limit}
r=\frac{r_++r_-}2+\frac\vareps2\rhat
~~,~~~~
t = \frac{\that}{2\vareps}
\ee
and dimensionally reducing over $\bS^1_\phi$, one has the nearly $AdS_2$ black hole metric and dilaton of JT gravity
\begin{align}
\begin{split}
\label{JTmetric}
ds^2 &= \frac{\ell^2}4\Big[ -(\rhat^2-1) \,d\that^2 \Big(1- \frac{\vareps\rhat}{2r_+}\Big) +\frac{d\rhat^2}{\rhat^2-1} \Big(1+\frac{\vareps\rhat}{2r_+}\Big) + O(\vareps^2) \Big]
\\[.2cm]
\Phi_{\rm\sst \!JT}^{~} &= \sqrt{g_{\phi\phi}} = r_+ + \hf\vareps(\rhat-1)\big) + O(\vareps^2)
\end{split}
\end{align}
together with Kaluza-Klein gauge fields from the metric and B-field
\begin{align}
\begin{split}
\label{AdS2potls}
A_\that &= -\frac{g_{t\phi}}{g_{\phi\phi}} = \frac{r_+ r_-}{r^2} \sim 1-\frac{\vareps\rhat}{r_+}+ O(\vareps^2)
\\[.2cm]
B_\that &= B_{t\phi} = r^2 - \sfc \sim  r_+(r_+ -\vareps)-\sfc + \vareps  r_+ \rhat + O(\vareps^2) ~.
\end{split}
\end{align}
The $AdS_2$ throat extends out to $\rhat\sim 1/\vareps$, at which point it opens out into the ambient $AdS_3$ geometry.  Note that in this throat, the size of the $\phi$ circle is approximately constant, and these two gauge fields are interchanged by T-duality on this circle.

One can now consider emission of quanta carrying the associated charges in this effective $AdS_2$ gravity theory, whose higher dimensional origin is KK momentum (charged under $A$) and string winding on $\bS^1_\phi$ (charged under $B$).  We thus expect a nonzero probability for the emission of winding strings from the black hole.

Of course, we need not work in the near-extremal limit; winding strings will be emitted from BTZ black holes of any mass and spin.  However the $AdS_2$ limit is conceptually useful, as it puts the winding strings on an equal footing with other charged quanta, such as pointlike quanta (unwound strings) carrying angular momentum, and we can use any of the standard techniques to compute the emission amplitude.

We thus first review the computation of particle emission, and then extend the discussion to winding strings.

%%%%%%%%%%%%%%%%%%%%%%%%%%%%%%%%
%%%%%%%%%%%%%%%%%%%%%%%%%%%%%%%%

\subsection{Emission amplitude: Particles}
\label{sec:pamplitude}

There are a variety of methods for calculating the emission amplitude for Hawking radiation.  Two closely related approaches that we will focus on are the tunneling and WKB methods (for a review, see~\rcite{Vanzo:2011wq}).  

In the tunneling approach, pioneered in~\rcite{Kraus:1994by,Kraus:1994fj,Parikh:1999mf}, one assumes that a particle-antiparticle pair is nucleated just inside the horizon, and computes the probability for the particle to tunnel across the horizon so that it can escape as a Hawking quantum.  The tunneling particle is treated as a shell of energy $E $ and angular momentum $L $.  As such, it starts its life inside a black hole of mass $M$ and angular momentum $J$ and leaves behind a black hole of mass $M-E $ and angular momentum $J-L$, and so the tunneling portion of its worldline starts just inside the initial black hole horizon and ends just outside the horizon of the final black hole.  

The calculation is typically done in Painlev\'e-Gullstrand coordinates, whose AdS version is described \eg\ in~\rcite{Hemming:2000as,Wu:2006pz}.  Starting from the coordinates $t,r,\phi$ of~\eqref{BTZmetric} or $\that,\rhat$ of~\eqref{JTmetric}, one reparametrizes the time coordinate to remove the coordinate singularity at the outer horizon, and the $\phi$ coordinate to account for frame dragging, through the transformations
\begin{align}
\label{paincoords}
dt = d\taub- \frac{\sqrt{1-F(r)}}{f(r)}\, dr
~~,~~~~
d\phi = d\varphi - \frac{J}{2r^2}\frac{\sqrt{1-F(r)}}{f(r)}\, dr
\end{align}
where one takes%
\footnote{There is freedom in the choice of $f_0(r)$; for instance, reference~\rcite{Hemming:2000as} makes the choice $f_0=1+r^2$ related to global AdS, while~\rcite{Wu:2006pz} included also the angular potential term from $f(r)$ as above.  This choice doesn't affect the near-horizon properties of the metric which determine the  tunneling rate.}
\be
F(r)=\frac{f(r)}{f_0(r)}
~~,~~~~
f_0(r)=r^2 + \frac{r_+r_-}{r^2}  ~.
\ee
The resulting metric is
\begin{align}
\begin{split}
\label{Paincoords}
ds^2 &= \ell^2\Big[ -f\,d\taub^2 + 2\sqrt{1-f/f_0}\, d\taub\,dr + \frac{dr^2}{f_0} + r^2\Big( d\varphi-\frac{J}{2r^2} d\taub\Big)^2 \,\Big]  ~.
\end{split}
\end{align}
In string theory, this coordinate transformation also generates a nonzero $B_{\taub r}$ and $B_{\varphi r}$ which we can (and do) remove by a suitable gauge transformation $\delta B=d\Lambda$; we then have $B_{\taub\varphi} = \ell^2\big( r^2 \tight- \sfc\big)$ as the only nonzero component of the $B$-field.

The tunneling amplitude is given by the imaginary part of the classical particle action~\eqref{Sptcl}. 
In the Hamiltonian path integral, one has
\be
\cS_\ptcl = \int \!d\xi\, \Big( p_\mu \dot x^\mu - \lapse\, \sfH(p,x) \Big)
\ee
where the Lagrange multiplier $\lapse=\sqrt{\gamma}/2$ enforces the Hamiltonian constraint~\eqref{Hamconst}.
We consider the tunneling portion of the trajectory near the horizon, which gives rise to an imaginary part of the action.  
Fixing the energy and angular momentum at infinity and accounting for the backreaction of the particle's conserved quantities $E,L$ on the ambient geometry, the trajectory runs between $r_{\rm in}=r_+(M,J)$ and $r_{\rm out}=r_+(M-E ,J-L)$.
The particle's Hamiltonian $\sfH$ vanishes, leaving the reduced action
\begin{align}
\begin{split}
\label{ImS}
\Im \cS_\ptcl &= \Im \Big[ \int_{r_{\rm in}}^{r_{\rm out}} \! dr \, p_r  \Big] 
=  \Im \Big[ \int_{r_{\rm in}}^{r_{\rm out}} \! \!dr \int_0^{p_r} \!\!dp'_r  \Big] 
\end{split}
\end{align}
where $p_r$ is the momentum conjugate to $r$; the reduced action integrals of the other conjugate pairs
\be
\int p_{\tau} d\tau+ \int p_\varphi d\varphi = E  \tau+ L  \varphi
\ee
are determined by the conserved quantities and only contribute to the real part of the action.

Solving the Hamiltonian constraint~\eqref{Hamconst} for the radial momentum, one finds a 
rather complicated function, but near the horizon it reduces to
\be
\label{prptcl}
p_r \sim \frac{E  r_+ - L  r_- }{(r_+^2-r_-^2)(r-r_+)}  ~.
\ee
In particular, this result is independent of the particle mass $\mu$, and the trajectory in the $r$-$\tau$ plane at the horizon becomes null.

Now note that we are working in a microcanonical ensemble of fixed total energy and angular momentum, and so the Hamiltonian constraints in the {\it spacetime} gravity theory relate the energy and angular momentum of the particle to the change in the energy and angular momentum of the black hole left behind~\rcite{Kraus:1994by,Kraus:1994fj,Parikh:1999mf}.  We can trade the integral over $p_r$ for an integral in $E ,L $, and then trade that for an integral over $\Mtil =M-E $ and $\Jtil =J-L $
\begin{align}
\begin{split}
dp_r &= \frac{\partial p_r}{\partial E }\bigg|_{\sst M,J} dE  + \frac{\partial p_r}{\partial L }\bigg|_{\sst M,J} dL   
~~\sim~ -\frac{\ell r_+}{r_+^2-r_-^2} \frac{d\Mtil  - \Omega_+ d\Jtil }{r-r_+}
\end{split}
\end{align}
We now reverse the order of integration, and do the $r$ integral first.
The pole along the integration contour is resolved by deforming it slightly into the lower half plane
and using the principle value prescription $\frac{1}{r-i\eps} = \cP( \frac1r) +i\pi\delta(r)$. 
The imaginary part of the integrand is thus
\be
-\frac{\pi \ell r_+}{r_+^2-r_-^2} \big(d\Mtil  - \Omega_+ d\Jtil \big) = - \frac{d\Mtil  - \Omega_+ d\Jtil }{2T} = -\half\,dS_\BH
\ee

One finds a remarkably simple and universal answer, that the imaginary part of the action is half the total change in the black hole entropy during the emission process
\begin{align}
\begin{split}
\label{tunnelamp}
\Im \cS_\ptcl &
 = \half \Big( S_\BH(M\tight-E ,J\tight-L) - S_\BH(M,J)\Big)  
 = \frac{2\pi\ell(r_{\rm out}-r_{\rm in})}{8G} 
 = \pi N (r_{\rm out}-r_{\rm in})
\end{split}
\end{align}
both for JT gravity~\rcite{Vagenas:2001sm,Aalsma:2018qwy} and for the BTZ black hole~\rcite{Vagenas:2001rm,Medved:2001ca,Liu:2005hj,Wu:2006pz}, and for that matter every other example that has been worked out~\rcite{Vanzo:2011wq}.
The same expression for the transition probability $\Gamma\sim \exp[\Delta S_\BH]$ even applies to the fragmentation of an $AdS_2$ black hole of mass $M$ and charge $Q$ into two smaller black holes of mass/charge $m,q$ and $M-m,Q-q$~\rcite{Maldacena:1998uz}.%
\footnote{A similar result is also expected to hold for the fragmentation of D1-D5-P (or F1-NS5-P) BTZ black holes at the singular locus in the CFT moduli space.  A calculation along the lines of the next subsection including fivebrane charge is relatively straightforward.}

The WKB approach describing pair creation at the horizon originated somewhat earlier in the work of Hartle and Hawking~\rcite{Hartle:1976tp}, and proceeds along rather similar lines.  One employs the fact that gravitational redshift makes the near-horizon evolution of the particle wavefunction have increasingly short wavelength, and so one can again employ WKB methods.  The leading term in the WKB expansion of the quantum amplitude%
\footnote{More precisely, the ratio of the real parts of the emission and absorption amplitudes.} 
is given by the solution to the Hamilton-Jacobi equation.
One follows the same steps as above to arrive at
\be
\label{ImSagain}
\Im \cS_\ptcl = \Im \Big[ \int_{r_{\rm in}}^{r_{\rm out}} \! dr \, p_r  \Big] 
= \Im \Big[\int_{r_{\rm in}}^{r_{\rm out}} \! dr \, \frac{E  r_+ - L  r_- }{(r_+^2-r_-^2)(r-r_+)}  \Big] 
\ee
but now, rather than incorporating the backreaction, one simply performs the integral over $r$ using the same contour deformation prescription.  The result is
\be
\label{ImSptclWKB}
\Im \cS_\ptcl = \frac{\pi}{\kappa_+}\big( E  - L \Omega_+\big) = -\half\frac{d \Mtil -\Omega_+ d \Jtil }{T}  =  -\half  {d S_\BH} 
\ee
in terms of the surface gravity $\kappa_+=2\pi T$ and angular potential $\Omega_+$ at the outer horizon.
Comparing to the result of the tunneling approach~\eqref{tunnelamp}, with the BTZ entropy
\be
\label{BTZent}
S_\BTZ = \frac{2\pi \ell r_+}{4G} = 2\pi N r_+ = 2\pi\left( \sqrt{N(\ell M+J)/2} + \sqrt{ N (\ell M-J)/2} \, \right)  
\ee
using the relations~\eqref{MJrpm}, one finds agreement in the limit ${E }/{\ell M},{L }/{J}\ll1$.

While these two calculations give the same result, their conceptual underpinnings are somewhat different.  In the WKB approach, one is computing Hawking pair production at the horizon; energy conservation is a consequence of the emitted Hawking quantum having an interior partner having the opposite energy and other conserved charges.  Along the particle worldline, the trajectory has the same $p_t,p_\phi$ along a path that is traveling forward in time outside the horizon and backwards in time inside the horizon; the opposite orientation reverses the sign of the charges.  The conserved charges of the black hole after emission are thus those of the initial black hole $M,J$, minus those of the emitted quantum $E ,L $, but the back-reaction of the emission process on the geometry is not incorporated in the leading order calculation.  The black hole after emission has energy $M-E $ and angular momentum $J-L $ not because the geometry has changed, but because the conserved quantities are the sum of those of the geometry $M,J$, plus those of the interior partner of the radiated quantum having charges $-E ,-L $.  The back-reaction of this interior partner on the black hole is treated as a subleading correction to the Hawking process.

In the tunneling approach, the usual assumption is that again a particle-antiparticle pair fluctuate into existence near the horizon, and then one computes the probability that the particle member of the pair tunnels out and becomes an on-shell quantum outside the black hole, drawing the energy to do so from the black hole itself.  But one need not ask where the Hawking quantum came from~-- the starting point is a spherical shell of energy placed just inside the horizon of the initial black hole of conserved charges $M,J$, which then tunnels out across the horizon to become a radiated Hawking quantum just outside the horizon of a black hole whose geometry has charges $M-E ,J-L $.  
Here we have in mind the fuzzball scenario as an alternative to pair creation near the horizon, where the black hole interior is not the vacuum BTZ geometry; then one can entertain the possibility that the quantum in question simply materializes from the ambient fuzz, and then tunnels out.  No partner anti-particle is ever involved, but the end result is the same as seen from the outside.  The difference lies in the fact that the pair production process leads to the information paradox, while emission from a fuzzball need not~-- the interior partner is supplanted by a changed state of the fuzzball, with which the emitted quantum is entangled.

In the tunneling analysis, energy conservation arises from solving the constraints of general relativity (including the contribution from the particle stress tensor).  Then the difference between the conserved charges of the black hole before and after the emission match those carried by the emitted quantum.  One then self-consistently computes the tunneling amplitude along a path that starts in one geometry and ends in the other, so the back-reaction of the emitted quantum on the geometry is fully taken into account.

\begin{figure}[ht]
\centering
\includegraphics[width=0.4\textwidth]{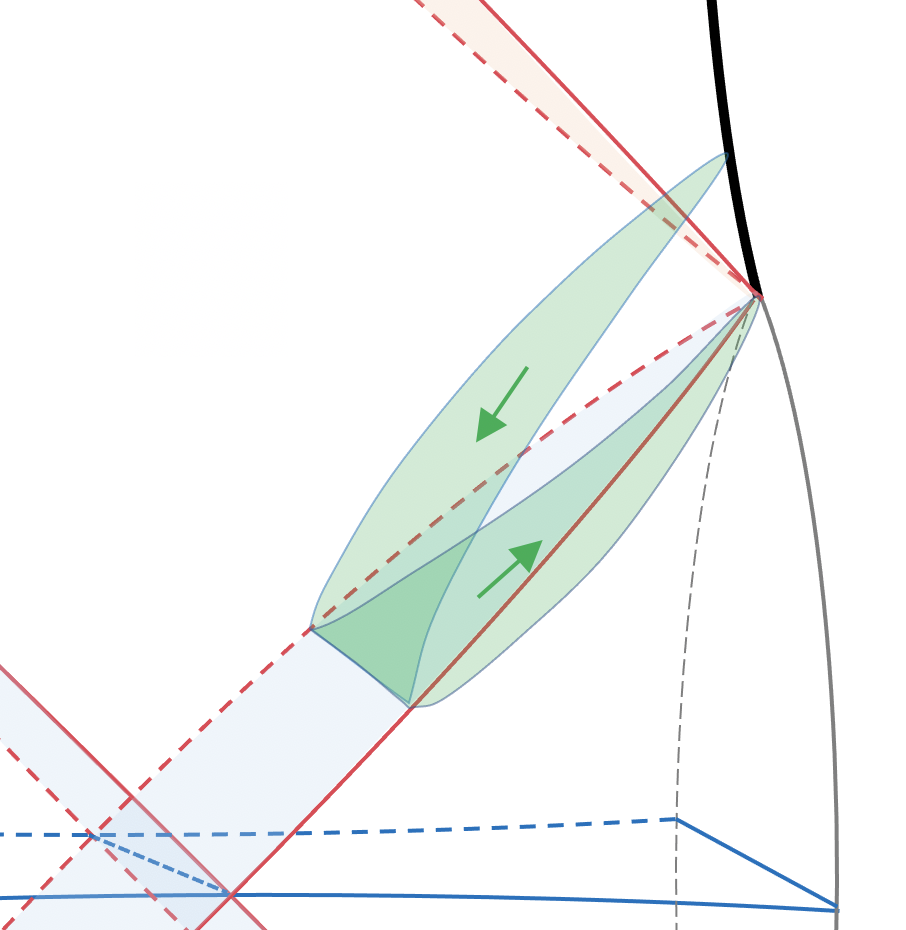}
\vskip .1cm
\caption{\it Pair creation of winding strings.  Arrows indicate the flow of proper time along the worldsheet.}
\label{fig:stringpair}
\end{figure}

The tunneling quantum is usually considered to be that of a particle in an S-wave state, but in our application we are also interested in the case where it is a string winding the $\phi$ circle, depicted in figure~\ref{fig:stringpair}.  We turn next to an analysis of this possibility.

\vskip .3cm
%
%%%%%%%%%%%%%%%%%%
%%%%%%%%%%%%%%%%%%
%

%%%%%%%%%%%%%%%%%%%%%%%%%%%%%%%%
%%%%%%%%%%%%%%%%%%%%%%%%%%%%%%%%

\subsection{Emission amplitude: Winding strings}
\label{sec:stramplitude}

The experience with the tunneling method in a wide variety of situations leads to a universal answer~-- the probability for a quantum to escape the black hole is 
\begin{align}
\begin{split}
\label{emitprob}
\Gamma \sim \exp\big[ {\Delta S_\BH} \big] 
\end{split}
\end{align}
In the reduction of $AdS_3$ string theory to JT gravity, both momentum and string winding on the $\phi$ circle reduce to gauge charges in JT gravity, coupled to the gauge potentials $A$ and $B$ of~\eqref{AdS2potls}.  The dimensionally reduced theory doesn't discriminate between these two kinds of charge, they are both simply gauge charges coupled to JT gravity.  One then expects that the emission probability for strings of energy $E $, $AdS_3$ angular momentum $L $, and winding $w$ is given at leading order in the near-extremal limit~\eqref{AdS2limit}-\eqref{AdS2potls} by an expression having the same form for both
\begin{align}
\begin{split}
\delta S_{\rm JT} = \frac 1T\Big[ \delta M - A(r_+) \,\delta J  - B(r_+) \,\delta n_1 \Big]  ~.
\end{split}
\end{align}
However, the three-dimensional origin of the charges is quite different~-- the ``charge'' $\delta J=L$ is $AdS_3$ angular momentum, and the charge $\delta n_1=w$ is string winding.  The potential $A(r_+)$ is then the angular potential $\Omega_+$ in $AdS_3$.  The $AdS_3$ origin of the potential $B(r_+)$ is rather different, however;
emission of string winding charge $\delta n_1$ changes the $AdS_3$ radius~\eqref{centchg}, and so we need to employ a formalism where the cosmological constant is one of the thermodynamic variables.

The inclusion of a variable cosmological constant in AdS black hole thermodynamics has been explored in~\rcite{Caldarelli:1999xj,Kastor:2009wy,Cvetic:2010jb}; the BTZ version specifically was analyzed in~\rcite{Wang:2006eb,Dolan:2010zz,Dolan:2010ha}.  The $AdS_3$ cosmological constant plays the role of a thermodynamic ``pressure'', and the associated conjugate thermodynamic ``volume'' is heuristically the amount of space excluded by the event horizon
\be
P = - \frac{\Lambda}{8\pi G}  
~~,~~~~
V = \ell^2 \pi ({r_+^2-\sfc}) 
\ee  
(recall that we are working in dimensionless coordinates referred to the AdS scale).  When $\sfc=0$ (as we will find below), $V$ is the naive area of a disk of size $r_+$.
The first law is thus modified to include a chemical potential $\Theta$ for changes in the cosmological constant
\be
\label{grandcanon}
dM = T  dS + \Omega_+\, dJ + \Theta \, d\Lambda  ~.
\ee
For BTZ black holes, we have in addition to~\eqref{MJrpm} and~\eqref{ImSptclWKB} the conjugate pair
\be
\Lambda = -\frac{1}{\ell^2}
~~,~~~~
\Theta = -\frac{\ell^2(r_+^2-\sfc)}{4G} 
 ~.
\ee
Indeed, in the extremal limit, we recover the potentials~\eqref{AdS2potls}.
It is useful to think of this additional term as related to a variation of $N=\ell/4G=kn_1$
\be
\Theta\, d\Lambda  = -\frac{r_+^2-\sfc}{\ell} \, dN \equiv \ThetaN \, dN  ~.
\ee
Note that this chemical potential is essentially the value of the $B$-field at the horizon.  Ordinarily, this quantity can be changed by a gauge transformation, and since this only changes propagation by an unphysical phase we don't care.  But in processes involving black holes, we do care, since the value of the gauge potential at the horizon is the chemical potential for charge emission, and thus has physical meaning.  In the end, we will choose the gauge for which the imaginary part of the action yields the change in entropy, and so is consistent with the thermodynamics.
In asymptotically flat spacetimes, it was argued in~\rcite{Ferrari:2016vcl} that one can fix the value of the potential by imposing that the action of a probe asymptotically be simply the mass of the object.  If we start from the asymptotically flat NS5-F1 geometry and adopt this prescription, then the decoupling limit indeed has $\sfc=0$.

A WKB/tunneling analysis again arrives at the expected result.  Consider the worldsheet path integral governed by the action~\eqref{Sstr}
with the metric~\eqref{Paincoords} and the B-field~\eqref{Bfield} (the latter transformed via~\eqref{paincoords}, with the resulting $B_{\taub r}$ and $B_{\varphi r}$ removed by gauge transformation).  Performing the Legendre transform to the Hamiltonian action
\begin{align}
\cS_\str = \int\Big( p_\mu \dot x^\mu - \sfN_0 \,\sfH - \sfN_1 \sfP \Big)
\end{align}
results in the $AdS_3$ contribution to the Hamiltonian constraint given by a rather complicated expression; the parts that depend on the radial momentum $\pbi_\rbi$ are
\begin{align}
\begin{split}
\sfH^{~}_{\rm AdS} &=  \frac{f(r)\,\pbi_\rbi^2}{2k} +\sqrt{1-f/f_0}\; \frac{\pbi_\rbi}{k}\Big(\tight-{\pbi_{\taub}} \tight+\frac{ r_+r_-\,\pbi_\varphib}{kr^2}\Big)
\\[.2cm]
&\hskip1cm
- \sqrt{1-f/f_0}\, B_{\taub \varphi}\, \pbi_\rbi \Big( \varphib'-\frac{r_+r_-}{r^2}\, \taub' \Big) 
+ \big( \pbi_\rbi \textrm{ independent} \big)
\end{split}
\end{align}
where $B_{\taub\varphib}=k(r^2-\sfc)$, and primes denote derivatives with respect to $\xi_1$.
Now set 
\be
\pbi_{\taub} = E_\str 
~~,~~~~
\pbi_\varphib = L _\str
~~,~~~~
\taub'=\rbi^{\,\prime}=0
~~,~~~~
\varphib'=w ~,
\ee
and solve the constraint $\sfH_{\!AdS}+\sfH_{other}=0$ for $\pbi_\rbi$ in the near-horizon limit ($\sfH_{other}$ being the contribution of the rest of the worldsheet dynamics).  Of the two solutions, we want the one which corresponds to the outgoing mode; it has a pole at the horizon, 
\be
\label{prstr}
\pbi_\rbi \sim \frac{   r_+ E_\str -   r_- L_\str +  kw\, r_+(r_+^2-\sfc)}{(r_+^2-r_-^2)(r-r_+)}  
\ee
and generalizes the result~\eqref{prptcl} to winding strings.  Once again, the pole term in $\pbi_\rbi$ does not care about the various oscillator excitations $N_L, N_R$ or other contributions $h,\bar h$ to the worldsheet conformal dimension; it only cares about the three conserved quantities $E_\str,L_\str,w$.  These other attributes of the string's state enter through the constraints $\sfH=\sfP=0$, which correlate them with the conserved charges as in~\eqref{EstrLstr}.
Following the same steps as for the particle analysis, we again find~\eqref{ImSptclWKB} for the WKB analysis, with the modification to the equation of state~\eqref{grandcanon}
\be 
\label{stringamp}
\Im \cS_\str = \frac{\pi}{\kappa_+}\Big( E_\str  -  \frac{r_-}{r_+} \, L_\str +  kw\,({r_+^2 - \sfc})\Big) 
= -\frac{1}{2T}\Big({d \Mtil -\Omega_+ d \Jtil  - \ThetaN \,d\Ntil }\Big)  = -\half  {d S_\BH} ~,
\ee
noting that $kw = \delta N =  \delta c/6$ is the change in the central charge~\eqref{centchg} in the vicinity of the black hole resulting from the emission of the string.

From the expression~\eqref{Scft} for the BTZ black hole entropy in terms of left and right conformal dimensions~\eqref{Lzero}, we can evaluate the change in entropy upon emitting a winding string in either of two ways.  In the first approach, we include the background energy~\eqref{deltaM} in the string energy~\eqref{EstrLstr}, write all quantities in Planck units and compute the change in BTZ entropy
\begin{align}
\begin{split}
\label{dSbtz}
\delta S_\BH &= 
2\pi \sqrt{\frac{(N\tight-kw)}{2}\Big[ \big((N\tight-kw)4G M\tight-E_\str\big) + (J\tight-L_\str)\Big] } 
\\[.2cm]
&\hskip 1cm
+ 2\pi \sqrt{\frac{(N\tight-kw)}{2}\Big[ \big((N\tight-kw)4G M\tight-E_\str\big) - (J\tight-L_\str)\Big] }
\\[.2cm]
&\hskip 2cm 
-2\pi \sqrt{\frac{N}{2}\Big[ \ell M + J\Big] } - 2\pi \sqrt{\frac{N}{2}\Big[ \ell M-J\Big] } ~~.
\end{split}
\end{align}
Expanding in the limit $E_\str \tight\ll \ell M, L_\str \tight\ll J, kw \tight\ll N$, one finds
\be
\label{dSc=0}
\delta S_\BH =  \frac{ 2\pi(r_+ E_\str-  r_- L_\str+ r_+^3 kw)}{r_+^2-r_-^2} 
=  \frac{ 2\pi(r_+ \delta E-  r_- \delta L)}{r_+^2-r_-^2} + \pi kw \,r_+
\ee
in agreement with~\eqref{stringamp} (with $\delta E,\delta L$ defined in~\eqref{deltaEL}), provided that $\sfc=0$.
This value was advocated in~\rcite{Ashok:2021ffx} for reasons related to the energy of the wound strings, and as discussed above, we also arrive at this choice using the prescription of~\rcite{Ferrari:2016vcl}.

Alternatively, we can instead use the expression~\eqref{BTZent} for the entropy, keeping $\ell M$ fixed instead of $4G M$ while we vary $\delta N=kw$, and measure the string's energy relative to threshold so that $\delta(\ell M)=\delta E$ as in~\eqref{deltaEL}.  One again arrives at~\eqref{dSc=0}.

The emission is predominantly in winding sector $w=1$, with the string unexcited \ie\ $h=\bar h=N_L= N_R=0$, and near threshold $s=0$.  The dominant contribution to the suppression of the emission of such winding strings near threshold comes from the term $\pi kw r_+$.  The black hole entropy is the CFT entropy~\eqref{Scft}; extracting some of the winding charge reduces the central charge $c=6N$ in the black hole by $\delta c=6kw$, and therefore the black hole entropy.  Since this entropy is larger than the winding string entropy by a factor $\sqrt{k}$, this loss of entropy can't be compensated by a corresponding excitation of the emitted fundamental string.  In other words, in computing the probability for F1 string emission, we have a sum over final states, and so gain a factor of $e^{S_\str(\delta E)}$.  But since the entropy of F1 strings is smaller than the BTZ entropy by a factor of order $\sqrt{k}$, one still finds that the overall probability of emission $e^{\delta S_\BH+S_\str(\delta E,\delta L)}$ is heavily suppressed.

The energy of the string is drawn from the black hole mass, and correspondingly the higher the energy the more the change in entropy, and the more suppressed the emission becomes.  The least costly are strings emitted in bound state trajectories, in their oscillator ground states $N_L= N_R=0$ and internally unexcited $h=\bar h=0$ (to the extent allowed by the level matching constraint $\sfP=0$), having $j\in\bR$ in the range~\rcite{Maldacena:2000hw}
\be
\frac{k+1}{2} > j > \half ~;
\ee
however these strings fall back into the black hole, and thus are simply part of its thermal atmosphere.  The continuum of holar wind states begins at the threshold $j=\half$ and ranges over $j=\half+is$, $s\in\bR$.  The situation is similar to the boiling off of a planetary atmosphere from the tail of its Boltzmann distribution (a phenomenon known as {\it Jeans escape}).  Similarly, here one has a thermal distribution of winding strings, and the tail of that distribution achieves escape velocity and wanders off to infinity.  These emitted strings will be largely in their ground state, and with very low radial momentum, since again the more excited it is or the more radial momentum it has, the more energy is taken away from internal excitations of the black hole, and the more the tunneling rate~\eqref{emitprob} will be suppressed.

%%%%%%%%%%%%%%%%%%%%%%%%%%%%%%%%
%%%%%%%%%%%%%%%%%%%%%%%%%%%%%%%%

\subsection{Extension to linear dilaton asymptotics}
\label{sec:lindil}

In the worldsheet formalism, the connection of $AdS_3$ holography to the parent fivebrane dynamics arises through a current-current deformation of the $\sltwo$ conformal field theory~\rcite{Giveon:1999zm,Giveon:2017nie}, which is also exactly solvable~\rcite{Hassan:1992gi,Giveon:1993ph}.%
\footnote{This deformation is often referred to as a ``$T\overline T$'' deformation, but that description is limited to a particular non-geometric corner of the moduli space of the spacetime CFT, where it is realized as a symmetric product orbifold.  The characterization of this deformation in the strong-coupling regions of the CFT moduli space where supergravity applies is less clear, though its effect on the long string spectrum takes the same form because the asymptotic structure of the long string Hilbert space is that of a Fock space (due to the vanishing dilaton), and hence a symmetric product.  The full CFT however likely does not have such a symmetric product structure.}
This deformation modifies the asymptotics of the background from that of $AdS_3$ to the linear dilaton asymptotics of decoupled fivebranes.  Its study gives us one of the few available tools to investigate NS5-brane holography, which is thought to be governed by a strongly-coupled noncritical string dynamics known as {\it little string theory}~\rcite{Maldacena:1996ya,Seiberg:1997zk,Dijkgraaf:1997ku} (for a review, see~\rcite{Kutasov:2001uf}).  The $AdS_3$ long string S-matrix is a limit of the winding sector S-matrix of little string theory holography~\rcite{Aharony:2004xn}.
The analysis of Hawking emission thus straightforwardly extends to the more general setting of little string theory.

The results above extend straightforwardly to the linear dilaton holography of little string theory.
In the spacetime geometry, one can consider the decoupling limit in two steps.  First, one can take the asymptotic value $g_s$ of the string coupling to zero; this decouples the fivebranes from the asymptotically flat region of spacetime, leaving the asymptotically linear dilaton throat of the $n_5$ fivebranes.  At a radial scale set by the number $n_1$ of strings in the background, the geometry rolls over to an $AdS_3$ regime at smaller radius, with the value of the dilaton saturating at $e^{2\Phi}\sim V_4 n_5/n_1$, where $V_4$ is the volume of $\cM$ in string units.  One thus has a smooth, weakly-coupled background when $n_1\gg n_5 V_4$.

\begin{figure}[ht]
\centering
\includegraphics[width=0.4\textwidth]{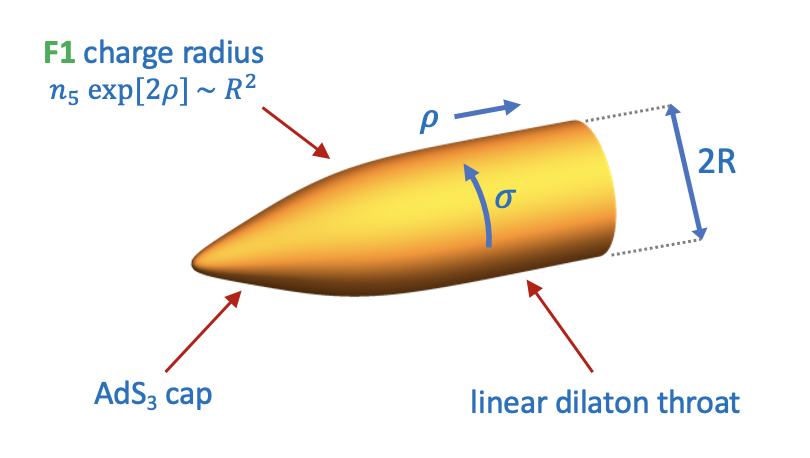}
\caption{\it Throat geometry of the NS5-F1 system.}
\label{fig:NS5F1geom}
\end{figure}

The $AdS_3$ decoupling limit arises when one further scales the radius $R$ of the circle to infinity, while scaling the energy above extremality to zero, keeping the product $ER$ fixed (the latter quantity becoming the scale dimension of the corresponding CFT state).  The spatial geometry is depicted in figure~\ref{fig:NS5F1geom}.  As $R$ is increased, the $AdS_3$ region persists out to larger and larger radius, until in the limit it comprises the entire spacetime.  Holography in $AdS_3$ thus arises as a particular limit of a particular superselection sector of NS5-brane holography.

\vskip .5cm
%
%%%%%%%%%%%%%%%%%%
%%%%%%%%%%%%%%%%%%
%

The details of the wound string emission process are only sensitive to the near-horizon geometry, and so are insensitive to whether we are in the AdS decoupling limit or just the fivebrane decoupling limit.  Of course, if the fivebranes are sufficiently excited, the horizon radius will lie in the linear dilaton region.  The result~\eqref{emitprob} has a natural extension to the fivebrane decoupling limit, where $S_\BH$ is now the black fivebrane entropy in the decoupling limit~\rcite{Maldacena:1996ya}.

%%%%%%%%%%%%%%%%%%%%%%%%%%%%%%%%%%%%%%%%%%%%%%%%%%%%%%%%%%%%%%%%
%%%%%%%%%%%%%%%%%%%%%%%%%%%%%%%%%%%%%%%%%%%%%%%%%%%%%%%%%%%%%%%%

\section{Unitarizing black hole radiance}
\label{sec:unitarity}

In radiation from a black hole, an emerging Hawking quantum joins a thermal atmosphere of such quanta.  In AdS, this thermal atmosphere tails off as a power of $r$, and all the radiation remains in the vicinity of the black hole.  The black hole is in equilibrium with its radiation, as previously emitted quanta fall back into the black hole, to be replaced by new Hawking quanta.  It is this tail of the wavefunction that one samples with boundary operators when coupling the CFT to a bath in order to model black hole decay.  Alternatively, when the AdS throat opens out onto flat space at some large radius, one is siphoning off this tail of the wavefunction, whose small amplitude at the top of the throat yields the source strength for radiation passing from the top of the throat into the asymptotically flat region.

It is sometimes said that information about the black hole state is ``available'' at the boundary in AdS/CFT.  Of course it is, in the exact non-perturbative theory~-- {\it after} the infalling data has fallen into the black hole, been processed through the phase space of black hole microstates with its mode fractionation, chaotic mixing, etc, and has subsequently reconstituted itself into Hawking quanta in the thermal atmosphere.  It takes time to set up that thermal atmosphere.  In the meantime, the data is inaccessible to effective field theory observables.  And the mechanism that maintains quantum coherence between the black hole interior and the thermal atmosphere remains mysterious.

We have seen that the same calculation that governs the emission probability of particles from BTZ black holes in NS flux compactifications, also governs the emission of winding fundamental strings, as depicted in figure~\ref{fig:stringpair}.  But whereas particles lie in normalizable $AdS_3$ wavefunctions that fall away to zero at the conformal boundary (the thermal atmosphere states discussed above), winding strings above a threshold lie in plane wave normalizable wavefunctions (when RR moduli vanish) that head out on timelike trajectories toward the conformal boundary.  In other words, a tiny fraction of the thermal atmosphere ``boils off'' and escapes the vicinity of the black hole~-- the ``holar wind'' of the title.  This stream of Hawking quanta leads to a black hole S-matrix in $AdS_3$, without the need for coupling the system to a bath, or connecting the AdS throat to an asymptotically flat region at some large radius.

The decay of the initial black hole via the Hawking process in effective string theory leads to the usual information paradox~-- the emitted strings are wholly entangled with their partners behind the horizon, and not at all entangled with the degrees of freedom making up the initial black hole.  For instance, their polarization states appear to be truly random and uncorrelated with the particular microstate of the black hole they emerge from.  They are thus incapable of carrying away information about the initial state, and small corrections to the Hawking process do not help~\rcite{Mathur:2009hf,Guo:2021blh}.  Large corrections to the Hawking process are required to restore unitarity to the quantum evolution.

The information problem is the issue of how the thermal atmosphere could possibly be coherently related to the state of the black hole interior if the latter cannot causally communicate with the exterior (or even if the communication is sufficiently weak).  In the classical limit, black holes are black because their phase space volume is infinite, and the information gets lost as it ergodically explores this phase space, and has zero amplitude to re-emerge.  At finite $N$, this volume is finite, and one expects that there is a tiny tail of the wavefunction outside the black hole, which is the thermal atmosphere.  But the question is how to make systematic corrections to the dynamics, so that the horizon is not a causal boundary that prevents the external tail of the wavefunction from being coherent with the interior and thus carry information about the black hole microstate.

If the degrees of freedom carrying the information of the black hole have a finite probability to appear near the horizon, then the small amplitude for them to tunnel out calculated above yields the amplitude for that information to be available to observations in the exterior.  But the quanta tunneling out need to be built from the entropic degrees of freedom of the interior; it is in this way that the exterior atmosphere is coherent with the interior (unless one wants to invoke some sort of non-local interactions, which is problematic for a variety of reasons~\rcite{Martinec:2022lsb}).

%%%%%%%%%%%%%%%%%%%%%%%%%%%%%%%%
%%%%%%%%%%%%%%%%%%%%%%%%%%%%%%%%

\subsection{The fuzzball scenario}
\label{sec:fuzzballs}

The small corrections theorem~\rcite{Mathur:2009hf,Guo:2021blh} says that $1/N$ corrections to the Hawking process cannot patch up the increasing disparity between the entanglement structure of the Hawking process and the entanglement structure of a unitary evaporation process.  A thermal atmosphere built by the Hawking pair creation process cannot be the starting point for a systematic computation of a unitary dynamics; an order one modification is necessary.  The wavefunction of the entropic degrees of freedom having support at the horizon, with Hawking quanta emerging from that part of the wavefunction, could be such an order one modification.

The {\it fuzzball paradigm} is a proposal for the needed large correction to the usual Hawking process 
(for a recent overview, see~\rcite{Bena:2022rna}).  It posits that the microstates comprising the black hole interior are complicated bound states of strings and branes, whose wavefunction extends out to the horizon.  In the fuzzball scenario, there is no individual, isolated quantum in the black hole interior almost entirely entangled with a given emitted Hawking quantum, and only weakly coupled to its environment; rather, the emitted quantum is entangled with the stringy stew which makes up the black hole interior and accounts for its microstates.  The radiation process is akin to that of thermal radiation from ordinary bodies, in which radiated quanta are entangled with the degrees of freedom making up the initial object rather than some new degrees of freedom that are added to it.
In the fuzzball paradigm, the radiation of a Hawking quantum of energy $E$ from a black hole of mass $M$ simply slightly de-excites the fuzzball state, into a sector of its state space with energy $M-E $ and corresponding entropy $S_\BH(M-E )$.

The surface of the fuzzball is not a horizon, in the sense that the fuzzball interior is supposed to be in causal contact with the exterior.  Our re-interpretation above of the tunneling calculation of~\rcite{Kraus:1994by,Kraus:1994fj,Parikh:1999mf} models radiation from a fuzzball as a process of tunneling out into the exterior geometry,
under the assumption that the fuzzball is well-approximated by the vacuum black hole geometry outside the horizon.  Then the tunneling calculation reproduces the Hawking calculation, because it doesn't care about the interior structure of the black hole state, or where the tunneling quantum came from; the only requirement is that it be ``available'' at the horizon to undergo the tunneling process.
The thermal atmosphere would then be no different than the tail of the wavefunction in a cavity containing a block of material with a large density of states.  The dominant support of the wavefunction is in the material, with only a small tail residing outside, and the two are coherently in contact.

The escaping quantum doesn't really see a {\it barrier} that it is tunneling across; a small transmission coefficient across a barrier works both ways, and would entail a small transmission coefficient {\it into} the black hole from outside; but the absorption coefficient of the black hole is essentially unity.  Rather, what we see from the tunneling calculation is that the emission probability is governed by the change in the entropy of the black hole, and so the density of initial and final states.  What is seen as ``tunneling'' in the effective dynamics, and a barrier to causal propagation, is simply that there is much more phase space available inside the black hole than outside, and emitting a quantum reduces the size of the available interior phase space.

We have focussed here on the well-developed picture of $AdS_3$ holography with NS fluxes, where the perturbative bulk dynamics of strings is well-understood.  As discussed above, one can think of this duality as a limiting case of little string theory dynamics.
The simplest example is realized when $n_5$ NS5-branes are bound to $n_1$ fundamental (F1) strings, so that $k=n_5$ and $N=n_5n_1$.   In the process of binding to the fivebranes, the fundamental strings fractionate into $n_5$ constituent {\it little strings}~\rcite{Seiberg:1997zk,Dijkgraaf:1997ku,Kutasov:2001uf,Martinec:2019wzw} whose tension scale $\alpha'_{\it little}=n_5\alpha'$ is also fractionated.  
The BTZ black hole entropy $S_\btz$~\eqref{BTZent} can be regarded as the entropy of little strings in a decoupling limit~\rcite{Maldacena:1996ya,Martinec:2019wzw}~-- $S_\btz$ is the entropy of little strings whose total winding charge $n_1n_5$ is $n_5$ times that of the fundamental strings.%
\footnote{In states near the vacuum, the background fivebranes are slightly separated onto their Coulomb branch, as one sees from string worldsheet constructions in such backgrounds~\rcite{Martinec:2017ztd,Martinec:2018nco,Martinec:2019wzw,Martinec:2020gkv,Martinec:2022okx}.  In this regime, the little strings are effectively massed up by the separation of the fivebrane source, and black holes are suppressed.  When fivebranes come together, the nonabelian little strings' tension becomes less than that of fundamental strings; the black hole phase arises through a deconfinement transition, much like other examples of holography.}

Our heuristic picture is that on occasion, an F1 string assembles itself near the horizon inside the bound state of fractionated ``quasistrings'' inside the black hole.  The calculation above determines the probability for such a reassembled F1 winding string in the black hole/brane bound state to leak out into the continuum of F1 states outside the black hole, with a result that agrees with the standard calculation involving pair creation.  What looks like tunneling across a light cone in the effective field theory, in the microscopic interpretation is a suppression of the transition due to a lack of available phase space rather than a causal barrier.
From outside, one cannot distinguish whether the emission of quanta from BTZ black holes involves vacuum pair creation, or instead causal emission from fuzzball states.

A similar picture of Hawking radiation arose in the BFSS matrix model~\rcite{Banks:1997hz,Klebanov:1997kv,Horowitz:1997fr,Li:1998ci}.  There, once again a black hole is a bound state of $N$ $Dp$-branes compactified on $\bT^p$, and Hawking radiation is the cleaving off of small clusters of these $Dp$-branes that are able to free themselves from the bound state and escape onto their Coulomb branch.  Simple virial estimates yield the horizon radius as the characteristic size of the brane bound state, and the characteristic properties of the escaping branes are those of Hawking radiation.  

In both of these examples, the radiation of charges carried by the black hole proceeds by the same sort of mechanism in operation at weak coupling.  At weak coupling, the radiation of D-brane charge from a D-brane bound state occurs by a freezing out of the open strings that are binding the emerging brane to the rest of the branes in the bound state.  The emerging charge comes directly from the branes that made the bound state to begin with.  The radiation of D1-branes from D1-D5 bound states is an example, S-dual to the F1-NS5 system we have been working with.%
\footnote{The F1-NS5 and D1-D5 effective string theories are valid descriptions in complementary parts of the CFT moduli space~\rcite{Seiberg:1999xz,Larsen:1999uk,Martinec:2022okx}.}  Incident D1-branes dissolve into D5-branes as fractional instantons, which are the weak-coupling description of little strings, and in the reverse process, the instanton strings assemble into a D1-brane that leaves the bound state.

The usual computation of Hawking radiation would have us believe that the way such brane charge emission occurs in the strongly coupled CFT is via a wholly different process, in which a brane-antibrane pair is nucleated at the horizon as in figure~\ref{fig:stringpair}, with the brane escaping to the asymptotic region.  In the remaining black hole, the original branes in the bound state remain at the black hole singularity sourcing the original antisymmetric tensor flux of the background and there is an antibrane in a negative energy state just inside the horizon.  The manifestation of the information paradox is that the branes returning to the asymptotic region are not related to the ones that made the initial black hole, but some other branes, generated by vacuum fluctuations.  For instance, the information about the polarization states of the incident strings making the intermediate BTZ black hole seems to be lost when those strings hit the BTZ singularity.
In the standard semi-classical picture, the polarization states of the emitted winding strings are random, and completely correlated with those of their partners behind the horizon rather than correlated with the state of the strings at the singularity.

In the fuzzball picture, that information is available at the horizon, and the emitted strings are scrambled versions of incident strings, which appear at the horizon and tunnel out to the exterior.  Their states are fully correlated with what the incident strings evolved into upon entering the black hole.

We have assumed for the purposes of the calculation above that we could use the vacuum black hole geometry until just inside the horizon, in order to evaluate that transition amplitude for Hawking radiation, but in the fuzzball paradigm, further into the interior lies a non-geometric fuzzball state.  If we were to assume that the vacuum geometry persists further into the interior to some extent, and the fuzzball degrees of freedom are confined within some radius $r_{\rm fuzz}<r_+$, the result would be inconsistent with the thermal properties of the radiation.  This latter possibility would lead to a much higher suppression of the emission amplitude, as the emerging quantum has to traverse a much longer non-classical trajectory through the semi-classical geometry from $r_{\rm fuzz}$ to $r_+$, and this would be inconsistent with the thermodynamics.%
\footnote{As discussed in~\rcite{Mathur:2012jk}, the fuzzball scenario does not preclude an approximation of vacuum freefall for the dynamics of sufficiently energetic infalling quanta.  This issue is a matter of the response function of the fuzz in various regimes of momentum transfer; the emission of Hawking quanta is a regime of very low radial momentum, while the infalling dynamics is high momentum.}

The analysis of section~\ref{sec:hawking} explains how a fuzzball could reproduce black hole thermodynamics in spite of the fact that the black hole microstates are not described by the vacuum geometry of gravitational effective field theory inside the horizon.  The reason is that very little of the Hawking process in the tunneling approach depends on this interior geometry~-- all one cares about is the geometry at the horizon, and that a quantum trying to leave the black hole sees an effective barrier that it needs to tunnel across.

%%%%%%%%%%%%%%%%%%%%%%%%%%%%%%%%
%%%%%%%%%%%%%%%%%%%%%%%%%%%%%%%%

\subsection{Other proposals}
\label{sec:others}

A number of other resolutions of the information paradox have been proposed in recent years.  Among them are the following:
\begin{enumerate}[(1)]
\item
Some unspecified process, possibly involving spacetime wormholes, transfers the entanglement from the partners of the Hawking quanta residing inside the black hole to degrees of freedom on the outside.  This scenario goes by the slogan ``ER=EPR''.  A growing region of the interior known as the {\it island} is secretly encoded in the state space of the exterior degrees of freedom.  One can thus in principle (even if it might be very difficult) manipulate the black hole interior by manipulating the radiation.  For an overview, see~\rcite{Almheiri:2020cfm}.
\item  
Exponentially small overlaps between (i) the state resulting from the Hawking process, and (ii) exponentially many perturbatively realized states in effective supergravity on smooth spatial hypersurfaces, are argued to lead to the needed large corrections (for a recent discussion, see~\rcite{Chakravarty:2020wdm,Balasubramanian:2022gmo,Balasubramanian:2022lnw}).  
The states of the effective theory are supposed to be an overcomplete basis of coherent states, and the failure to take into account their overlaps leads to an incorrect assessment of the degree of entanglement of the radiated quanta.
\item
The effective supergravity/string theory states generated by the evaporation process live in a Hilbert space whose dimension can be much larger than that of the black hole microstates at late time, and so to match their properties the former must be projected onto the latter~\rcite{Akers:2022qdl}.  The map of the state generated by the Hawking process of effective field theory to the Hilbert space of the exact theory is said to be {\it non-isometric}~\rcite{Verlinde:2012cy,Marolf:2013dba}.
\end{enumerate}

%%%%%%%%%%%%%%

Effective field theory dynamics for black holes is something of a Ponzi scheme, that seeks to hide from the entanglement-regulating authorities the ever-growing entanglement between black hole and radiation generated by the Hawking process, and the missing entanglement between early and late Hawking quanta.  

\begin{figure}[ht]
\centering
    \includegraphics[width=.35\textwidth]{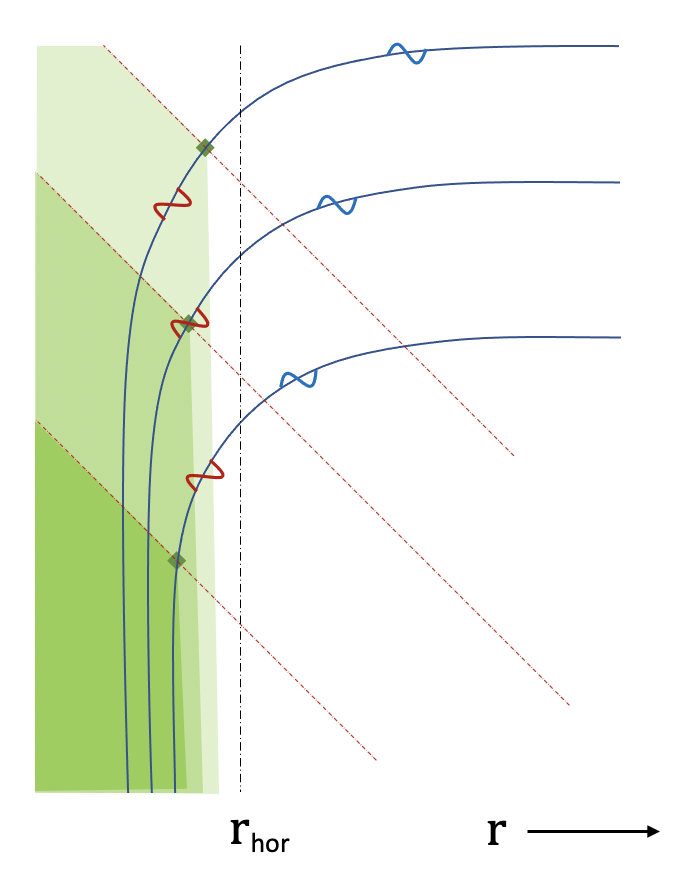}
\caption{\it 
The partner (red) of a Hawking quantum (blue) is transferred to the island.  As time evolves, these quanta both travel outgoing null trajectories while the island (shaded green) moves closer to the horizon.  In the earliest time slice, the interior partner quantum is outside the island (the dark green shaded region).  The middle time slice depicts the moment of the transition.  In the latest time slice, the Hawking partner lies in the island (the lightest green-shaded region), and is declared to belong to the radiation state space.
}
\label{fig:island-xfer}
\end{figure}

In option (1), the degrees of freedom entangled with the Hawking radiation are sequestered on an ``island'' in the black hole interior, and when the missing entanglement can no longer be ignored, it is declared not to be missing after all, by cooking the books~-- simply declaring that the island is part of the radiation, without specifying a mechanism for how it got there.
The evolution of the island is depicted in figure~\ref{fig:island-xfer}.
This accounting trick runs the danger of introducing non-local interactions into the dynamics that ought to be observable from outside the black hole~\rcite{Martinec:2022lsb} when one tries to specify a mechanism for this transfer of a portion of the black hole interior to the radiation.

%%%%%%%%%%%%%%%%%%
%%%%%%%%%%%%%%%%%%

In the context of the holar wind, the claim would be that the state of the black hole interior, including a large collection of Hawking pair partners residing on the island, is somehow encoded in the entanglement structure of the radiated winding strings at late times.
A consequence is that one can manipulate the state of the interior by manipulating the radiation.  It is hard to see how a weakly coupled collection of winding strings is going to reproduce the fast scrambling dynamics of the black hole interior.  If the island forms and some additional strings are tossed into the black hole, how are they supposed to strongly interact with a collection of early-radiated strings that are arbitrarily far away when the additional strings enter the black hole?  It seems that strong, arbitrarily nonlocal interactions are needed.

%%%%%%%%%%%%%%

Option (2) leads to the idea that a state of the effective field theory obtained by adding an additional Hawking partner behind the horizon can be expressed as a linear combination of states without such a quantum.  This is similar in some respects to the fuzzball proposal, in that in the exact theory, the interior partner of the radiated Hawking quantum is something of a fiction; the black hole interior is actually a state in the $e^{S_\BH(M-E )}$-dimensional black hole Hilbert space after emission, rather than in a state in the $e^{S_\BH(M)+\log 2}$ dimensional Hilbert space that results from the Hawking process, in which the radiated quantum is not entangled with the degrees of freedom accounting for the $e^{S_\BH(M)}$ initial states.  Therefore, the added interior excitation has to be able to be expressed in terms of a superposition of microstates in a Hilbert space of exponentially smaller size.

The distinction between this scenario and the fuzzball picture is an assumption that effective field theory is a valid description of the black hole interior, with the consequence that the degrees of freedom that the radiated Hawking quantum is entangled with are approximately decoupled from the rest of the black hole interior in the exact description, after rewriting it in terms of the exact microstates.  One has to be saying more than just that the interior partner to the radiated Hawking quantum is whatever it is entangled with in the exact theory~\rcite{Guo:2021blh}.  In order for effective field theory pair creation to be a good approximation to the dynamics, whatever the emitted member of the pair is entangled with has to be approximately decoupled from its environment (\ie\ one treats it as a quasiparticle or quasi-string).  This seems at odds with the notion that the degrees of freedom in the black hole interior are undergoing fast scrambling at the rate
\be
t_{\rm scr} \sim \frac{\beta_\BH}{2\pi} \log\big(S_\BH-S_0\big)  
\ee
where $S_0$ is the ground state entropy with the same charges.  It is hard to see how, even if one could manage to realize a geometric effective description of the interior partner just after its creation, as some complicated superposition of orthonormal basis states, that it wouldn't immediately fall apart as a result of the chaotic mixing going on in the interior state space.  At best, such a quasiparticle or quasi-string excitation would only be expected to persist for of order the scrambling time before it disappears, but then it seems that effective field theory has broken down for a description of the interior.
By way of contrast, in the fuzzball scenario the radiated quanta are entangled with the degrees of freedom of the fuzzball, which one expects are indeed all undergoing fast scrambling; one does not expect that the degrees of freedom with which the radiated quanta are entangled are somehow sequestered from the fast scrambling dynamics of the black hole interior.  

%%%%%%%%%%%%%%

Option (3) is a somewhat different proposal~-- that the effective field theory state space is not a subspace of the exact theory, but rather a much larger space that can only be projected onto the state space of the exact theory.  The states of the effective field theory are still orthogonal, and are not obtained by any sort of coarse-graining of the exact theory.  Since there is no isometry between the degrees of freedom of the effective field theory (or its perturbative string extension) and those of the exact theory, there is nothing to explain~-- the effective field theory description in terms of the Hawking process has nothing to do with the exact dynamics, and is in no sense an approximation of it.  It is argued that the projection operator that maps the EFT state onto that of the exact theory can be arranged to have the same statistical properties as the exact theory for a large class of EFT observables measuring the black hole exterior (at least, the model exhibited in~\rcite{Akers:2022qdl} has this character) .  This property is in line with the Eigenstate Thermalization Hypothesis~-- that generic/simple observations can't tell the difference between a pure state and a density matrix just by looking at a sample (substantially less than half) of the radiation.
One obtains the Page curve for the entanglement of the EFT Hawking radiation with the black hole; but this feature is a property of the {\it projection operator}, induced from the Page curve of the underlying exact theory, rather than being a feature of the effective dynamics itself.

One question to ask is whether any feature of this proposal would change if the fuzzball proposal were the correct exact description of the bulk dynamics.  Then there is no Hawking process going on at the horizon, and no entangled partners of the radiated Hawking quanta in the black hole interior.  But the claim would be that one can project the state resulting from the Hawking process onto the fuzzball state, and not be able to tell the difference for a wide variety of external measurements.  The EFT state (and particular the notion of interior partners of the emitted quanta) would be a complete fiction, and it is not clear what we would learn by considering it.  One aspect of the toy model of~\rcite{Akers:2022qdl} which might address this question is the construction of observables in the toy exact theory which approximate those of the toy model of EFT; if such a construction persists in actual examples of holography, it might lead to an understanding of how the interior dynamics might approximate that of geometric effective field theory for some purposes.

%%%%%%%%%%%%%%%%%%%%%%%%%%%%%%%%
%%%%%%%%%%%%%%%%%%%%%%%%%%%%%%%%

\subsection{Final remarks}
\label{sec:lastword}

Black hole interiors in the fuzzball scenario are a roiling, chaotic stew of branes.  The horizon is expected to be a sort of phase boundary, with these underlying branes in a deconfined state on the inside and a confined state on the outside.  The analytic continuation of the exterior vacuum geometry to the black hole interior would then amount to the analytic continuation of the exterior confined phase across a phase boundary, where it holds no validity (much like the analytic continuation of the Schwarzschild vacuum that describes the exterior of a neutron star is very much the wrong description of its interior).  Nevertheless, one might think of calculations using a geometric picture of the black hole interior, or Euclidean methods such as replica wormholes, as having some degree of validity, much as the continuation of a path integral to a complex saddle accurately approximates the path integral even if the saddle point isn't a real thing.  Indeed, classical gravity is a collective field theory that encapsulates black hole thermodynamics, and the classical vacuum black hole solutions (and especially their Euclidean continuations) might be considered such saddles which one fruitfully uses to compute ensemble averages of observables.  The collective field theory never keeps track of the microstates, and so asking it to keep track of the flow of entanglement and quantum information may be a bridge too far.  In this regard, the sorts of computations which have been used to reproduce the Page curve~\rcite{Penington:2019npb,Almheiri:2019qdq,Penington:2019kki,Chakravarty:2020wdm} should perhaps be thought of not as not exhibiting the microscopic mechanism of unitary black hole evaporation so much as capturing features of the ensemble's evolution.  For instance, the statement that ``the island is part of the radiation'' might simply be a sophisticated accounting device that keeps track of the flow of entanglement and quantum information, without reflecting the underlying microphysics responsible for it.  Euclidean methods in particular compute ensemble average quantities but don't reflect individual Lorentzian microstates or their evolution.  A sophisticated example of this phenomenon is the computation of the large charge asymptotics of BPS partition functions using a detailed analysis of the semiclassical expansion of Euclidean supergravity (see for example~\rcite{Dabholkar:2012zz,Cabo-Bizet:2018ehj,Iliesiu:2022kny}).  These computations aren't counting individual microstates, but the gravitational saddle points count the ensemble of them nonetheless.  In the present context, modular properties of 2d CFT partition functions relate the asymptotic density of states to the identity block \ie\ Chern-Simons gravity, which itself has almost no states.  Gravity isn't itself exhibiting the microstructure or its dynamics; but it sees the average properties of microstructure and keeps track of its collective behavior.  It may be that similarly, the island story and related ideas will eventually be regarded in the same light.

%%%%%%%%%%%%%

\vspace{1mm}

\section*{Acknowledgements}
%%%%%%%%%%%%%

%We thank 
%
%
The work of EJM is supported in part by DOE grant DE-SC0009924. 
Eugene N. Parker, to whom this paper is dedicated, was my longtime colleague on the UChicago faculty who in 1957 proposed the existence of the solar wind.  The proposal was met with strong opposition, and the initial referees rejected the paper.  It is hoped that the analysis above meets with a more receptive audience.

%\newpage
\vskip 3cm

\bibliographystyle{JHEP}      

\bibliography{fivebranes}

%%%%%%%%%%%%%%%%%%%%%%%%%%%%%%%%%%%%%%
%%%%%%%%%%%%%%%%%%%%%%%%%%%%%%%%%%%%%%

\end{document}

%%%%%%%%%%%%%%%%%%%%%%%%%%%%%%%%%%%%%%
%%%%%%%%%%%%%%%%%%%%%%%%%%%%%%%%%%%%%%